\PassOptionsToPackage{unicode}{hyperref}
\PassOptionsToPackage{hyphens}{url}
\documentclass[
]{article}
\usepackage{amsmath,amssymb}
\usepackage{iftex}
\ifPDFTeX
  \usepackage[T1]{fontenc}
  \usepackage[utf8]{inputenc}
  \usepackage{textcomp} 
\else 
  \usepackage{unicode-math} 
  \defaultfontfeatures{Scale=MatchLowercase}
  \defaultfontfeatures[\rmfamily]{Ligatures=TeX,Scale=1}
\fi
\usepackage{lmodern}
\ifPDFTeX\else
\fi
\IfFileExists{upquote.sty}{\usepackage{upquote}}{}
\IfFileExists{microtype.sty}{
  \usepackage[]{microtype}
  \UseMicrotypeSet[protrusion]{basicmath} 
}{}
\makeatletter
\@ifundefined{KOMAClassName}{
  \IfFileExists{parskip.sty}{%
    \usepackage{parskip}
  }{
    \setlength{\parindent}{0pt}
    \setlength{\parskip}{6pt plus 2pt minus 1pt}}
}{
  \KOMAoptions{parskip=half}}
\makeatother
\usepackage{xcolor}
\usepackage[margin=1in]{geometry}
\usepackage{color}
\usepackage{fancyvrb}

\DefineVerbatimEnvironment{Highlighting}{Verbatim}{commandchars=\\\{\}}
\newenvironment{Shaded}{}{}

\newcommand{\AttributeTok}[1]{\textcolor[rgb]{0.49,0.56,0.16}{#1}}

\newcommand{\CommentTok}[1]{\textcolor[rgb]{0.38,0.63,0.69}{\textit{#1}}}

\newcommand{\DecValTok}[1]{\textcolor[rgb]{0.25,0.63,0.44}{#1}}

\newcommand{\FloatTok}[1]{\textcolor[rgb]{0.25,0.63,0.44}{#1}}
\newcommand{\FunctionTok}[1]{\textcolor[rgb]{0.02,0.16,0.49}{#1}}

\newcommand{\KeywordTok}[1]{\textcolor[rgb]{0.00,0.44,0.13}{\textbf{#1}}}

\usepackage{longtable,booktabs,array}
\usepackage{calc} 
\usepackage{etoolbox}
\makeatletter
\patchcmd\longtable{\par}{\if@noskipsec\mbox{}\fi\par}{}{}
\makeatother
\IfFileExists{footnotehyper.sty}{\usepackage{footnotehyper}}{\usepackage{footnote}}
\makesavenoteenv{longtable}
\usepackage{graphicx}
\makeatletter
\def\maxwidth{\ifdim\Gin@nat@width>\linewidth\linewidth\else\Gin@nat@width\fi}
\def\maxheight{\ifdim\Gin@nat@height>\textheight\textheight\else\Gin@nat@height\fi}
\makeatother
\setkeys{Gin}{width=\maxwidth,height=\maxheight,keepaspectratio}
\makeatletter
\def\fps@figure{htbp}
\makeatother
\providecommand{\tightlist}{%
  \setlength{\itemsep}{0pt}\setlength{\parskip}{0pt}}
\newlength{\cslhangindent}
\newlength{\csllabelwidth}
\newlength{\cslentryspacingunit} 
\newenvironment{CSLReferences}[2] 
 {
  \setlength{\parindent}{0pt}
  \ifodd #1
  \let\oldpar\par
  \def\par{\hangindent=\cslhangindent\oldpar}
  \fi
  \setlength{\parskip}{#2\cslentryspacingunit}
 }%
 {}
\usepackage{calc}

\makeatletter
\@ifpackageloaded{subfig}{}{\usepackage{subfig}}
\@ifpackageloaded{caption}{}{\usepackage{caption}}
\AtBeginDocument{%

}
\AtBeginDocument{%

}
\newcounter{pandoccrossref@subfigures@footnote@counter}
{\end{figure}%
\addtocounter{footnote}{-\value{pandoccrossref@subfigures@footnote@counter}}
\@for\f:=\global@pandoccrossref@subfigures@footnotes\do{\stepcounter{footnote}\footnotetext{\f}}%
\gdef\global@pandoccrossref@subfigures@footnotes{}}
\@ifpackageloaded{float}{}{\usepackage{float}}
\floatstyle{ruled}
\@ifundefined{c@chapter}{\newfloat{codelisting}{h}{lop}}{\newfloat{codelisting}{h}{lop}[chapter]}
\floatname{codelisting}{Listing}

\makeatother
\ifLuaTeX
  \usepackage{selnolig}  
\fi
\IfFileExists{bookmark.sty}{\usepackage{bookmark}}{\usepackage{hyperref}}
\IfFileExists{xurl.sty}{\usepackage{xurl}}{} 
\hypersetup{
  pdftitle={The Battle of the Water Futures},
  pdfauthor={Dennis Zanutto*; Christos Michalopoulos; Lydia Tsiami; André Artelt; Jasmin Brandt; Demetrios Eliades; Stelios Vrachimis; Stefano Alvisi; Valentina Marsili; Filippo Mazzoni; Panagiotis Samartzis; Barbara Hammer; Phoebe Koundouri; Marios Polycarpou; Dragan Savić},
  pdfkeywords={water, future, conflict, climate},
  hidelinks,
  pdfcreator={LaTeX via pandoc}}

\title{The Battle of the Water Futures}
\author{Dennis Zanutto* \and Christos Michalopoulos \and Lydia
Tsiami \and André Artelt \and Jasmin Brandt \and Demetrios
Eliades \and Stelios Vrachimis \and Stefano Alvisi \and Valentina
Marsili \and Filippo Mazzoni \and Panagiotis Samartzis \and Barbara
Hammer \and Phoebe Koundouri \and Marios Polycarpou \and Dragan Savić}
\date{November 2025}

\begin{document}
\maketitle

\hypertarget{executive-summary}{%
\section*{Executive Summary}\label{executive-summary}}

This competition challenges participants to confront one of water
distribution systems's most pressing problems: adaptive planning under
deep uncertainty. Participants will develop a 75-year national
masterplan for a hypothetical Dutch water grid that must balance four
competing objectives: financial sustainability, service reliability,
environmental impact, and equitable access. Unlike traditional planning
exercises, this competition requires strategic decisions across three
25-year stages while navigating uncertain climate, energy, and economic
scenarios. Success demands developing an adaptive masterplan that blends
policy decisions and infrastructure interventions, gets updated based on
evidence accumulated across the implementation stages, and preserves
flexibility to avoid locking into suboptimal pathways. This document
defines the competition framework; a subsequent report will present
participant solutions and outcomes.

\tableofcontents

\newpage

\hypertarget{introduction}{%
\section{Introduction}\label{introduction}}

\hypertarget{disclaimer}{%
\subsection{Disclaimer}\label{disclaimer}}

The networks and the problem presented in this competition are synthetic
and, despite being based on the Netherlands, do not represent any
real-world situation. While we utilize real data from trusted sources
where possible (e.g., CBS\footnote{CBS (Centraal Bureau voor de
  Statistiek) is the Dutch national statistics agency:
  https://www.cbs.nl} for municipal data, KNMI\footnote{KNMI (Koninklijk
  Nederlands Meteorologisch Instituut) is the Dutch meteorological
  institute: https://www.knmi.nl} for weather observations,
Vewin\footnote{Vewin is the association of Dutch drinking water
  companies: https://www.vewin.nl} for statistics about water
utilities), other aspects rely on assumptions based on expert knowledge.

We are committed to the model's integrity and, based on competitor
feedback, we will continuously test and refine the model (e.g.,
balancing electricity and infrastructure costs) to ensure it remains
physically sound and balanced. If you identify any significant
inconsistencies, please inform us and we will review them promptly.

All versions are archived on Zenodo, with the latest version
establishing the official standard (Zanutto et al. 2025). An open-source
evaluator will also be released in the WaterBenchmarkHub (Artelt et al.
2025), allowing competitors to test their solutions before the final
submission.

\hypertarget{background-and-motivation}{%
\subsection{Background and motivation}\label{background-and-motivation}}

Water distribution systems (WDS) grow organically with cities and are
therefore inherently subject to deep uncertainties. Strategic WDS
planning focuses on the long-term design of the primary supply system --
the national grid. While secondary and tertiary network components can
typically be selected using established engineering principles, planning
the primary and supply network at a regional level requires a different
approach, one that integrates policy considerations, standard
engineering practice, and substantial investment. This part of the
system is extremely critical. Failures at this level can propagate and
lead to loss of service for thousands of people and businesses. Beyond
system reliability and deep uncertainties, the complex governance
requirements, such as the coordination between national, regional, and
local administrations, present a major obstacle to integrated
policy-making and long-term planning of these systems.

Over the past decade, the field has steadily refined and built upon
several approaches that address WDS planning under uncertainties. To
begin, the introduction of staged design, where interventions are
modelled in phases rather than in a single batch representing the target
network, provided a tool to represent masterplans with more realism
(e.g., Creaco, Franchini, and Walski 2014). Next, this modelling
framework was extended to include uncertainty, either through a robust
approach that accounts for it (e.g., Creaco, Franchini, and Walski
2015), or through a flexible approach that develops a single plan with
multiple pathways. In its simplest form, this takes the shape of a
decision tree, where only one branch is implemented based on how
uncertainties unfold (e.g., Basupi and Kapelan 2015).

An inherent limitation of robust and flexible plans is that they require
accounting for all possible future system interventions and scenarios at
the initial planning stage. Therefore, when uncertainties move away from
the expected values or new opportunities emerge (i.e., under deep
uncertainties), these approaches fall short, and the water utilities
must adapt to the changes in their environment.

An adaptive plan would typically feature at least one of two components.
The first component involves periodically re-evaluating and adjusting
the plan as uncertainties evolve, allowing new opportunities and
information to be incorporated. For example, see Beh, Maier, and Dandy
(2015) on water supply systems, and Skerker et al. (2023) on ``the value
of learning'' in flexible planning for water reservoirs. The second is
the ability to shift between pathways and avoid locking up to one (see
dynamic adaptive policy pathways for the Rhine Delta, Haasnoot et al.
2013).

Regardless of definitions and classifications, the field also lacks
coherence in benchmark definition, as none of these approaches have been
systematically tested against each other. At best, works introducing
different methodologies may have used the same benchmark networks (e.g.,
Anytown or New York Tunnels), but the overall problem descriptions---and
thus the assumptions and scenarios surrounding these networks---differ
substantially.

Therefore, in the context of this Battle of the Water Futures (BWF),
\textbf{the participants' goal is to apply innovative methodologies to
develop a masterplan for a national water grid under deep
uncertainties}. The organisers, in turn, seek to establish a benchmark
that advances decision-making practices for water distribution systems
under deep uncertainty, while laying the ground for future benchmarking
through a highly customizable, open-source framework.

\hypertarget{ambition}{%
\subsection{Ambition}\label{ambition}}

Participants must develop masterplans for a national water grid with
water utilities operating across different provinces. However, the BWF
presents a level of detail and computational complexity such that the
featured national grid optimisation problem cannot be solved through
brute force optimisation. Moreover, unlike traditional optimisation
problems with specified scenarios, here, only a description of the
environment dynamics and a handful of expert-driven scenarios for key
drivers (e.g., population) are provided. Competitors must embrace
uncertainty and decide for themselves what to model and how to do so,
reflecting the reality that different teams would make different
modelling choices when facing deep uncertainty.

The decision to place participants in a context of deep uncertainty is
intentional. The BWF will unfold over three competition stages, each
spanning 25 years, to realistically emulate the long-term challenges
faced by planners. This means that some parameters are uncertain and
their uncertainty bounds evolution will be provided only through
approximate expert-based scenarios (e.g., population projections vary
between minimum and maximum values, though actual outcomes may fall
outside these bounds). At the end of each competition stage, past
observations will be made available, enabling participants to implement
adaptive strategies and learn from accumulated evidence. Similarly, the
competition ranking process has been deliberately left undisclosed; only
the system requirements, namely, the metrics under which the system is
evaluated, are outlined in this document. This approach is more akin to
realistic situations and compels participants to identify their own
compromise solutions within this many-objective context.

\hypertarget{problem-description}{%
\section{Problem Description}\label{problem-description}}

The Battle of the Water Futures (BWF) problem focuses on a hypothetical
national drinking water grid of the Netherlands. The system structure
(demand nodes, sources, and links), available interventions (e.g.,
pipes, treatment plants), and exogenous drivers (e.g., population,
climate) are based on publicly available data about the Dutch context,
supplemented by assumptions or insights from other regions as needed.

The main objective of the BWF is to keep the system operational until
the end of the century, thereby preventing bankruptcy while fulfilling
the required service requirements in each water utility. Secondary
objectives include reducing emissions toward climate neutrality and
ensuring fairness across regions and generations. To achieve these
goals, the country's water utilities come together to develop a national
masterplan, which they agree to update every 25 years.

The masterplan specifies the interventions to be applied each year, both
at the national level (e.g., connecting independent water utilities) and
at the individual utility level (e.g., opening new sources). The
masterplan does not need to include the replacement of ageing components
(pipes and pumps), as these are automatically replaced when they reach
their end of life. However, to manage this uncertainty, utilities can
either schedule replacements in advance or allow components to age
naturally and account for replacement costs in the budget. The
masterplan also defines the national and regional policies and whether
they are revised during the planning period (note: while interventions
must be specified annually, policies can be set initially and remain as
they are until amended in the plan).

To prepare the masterplan, information on the current system status is
made available, including network models, prices, demands, population,
and other relevant parameters. Future projections of key drivers (e.g.,
population growth, climate trends) are also provided to complement
system knowledge. However, these scenarios are based on current expert
knowledge and should not be treated as precise forecasts as these
variables are generally deep uncertainties of the system.

Examples of deep uncertainties include:

\begin{itemize}
\tightlist
\item
  Desalination plant construction time: Planners are confident it can be
  built in 5 years, but acknowledge that delays could extend completion
  to 10 years.
\item
  Population evolution: Population is expected to follow the mean values
  or remain within the upper and lower bounds described by the national
  statistics agency. However, major external events, such as conflicts
  or mass migration driven by climate catastrophes, could push evolution
  beyond these ranges.
\end{itemize}

Behind the scenes, expert opinions are combined with dynamic models to
generate the synthetic time series that drive system evolution. At each
timestep, EPANET hydraulic simulations verify the network's physical
capability to transfer water and meet demands under the proposed
interventions and operating conditions.

Once the masterplan is defined, utilities begin following it, operating
and maintaining their systems according to its specifications. Around
them, the world evolves and uncertainties unfold. After 25 years,
utilities reconvene to update their plan. All system information is
updated, and the performance over the past 25 years along with observed
variables can be used to revise or completely redraw the masterplan.
This cycle---plan, execute, observe, adapt---repeats for three rounds,
testing the utilities' ability to manage both short-term operations and
long-term strategic planning under deep uncertainty.

\hypertarget{system-description}{%
\subsection{System Description}\label{system-description}}

This chapter describes the physical components that constitute the water
transport network in the BWF competition. The key entities are
municipalities, water sources, pumping stations, and pipes and each of
the following section will detail their characteristics, operational
constraints, and how they evolve over time. The physical feasibility of
any network configuration is verified through hydraulic simulation in
EPANET. External factors affecting the system, such as climate
conditions and the electricity grid, are addressed separately in
Section~\ref{sec:ext-drivers}.

\hypertarget{water-utilities}{%
\subsubsection{Water Utilities}\label{water-utilities}}

Within the BWF, Water Utilities (WUs) act as the entities responsible
for managing the assets and delivering drinking water across their
assigned provinces.

WUs must choose the policies to apply to their own system (e.g, setting
the budget for non-revenue water reduction) and decide the interventions
to take on their system (e.g., replacing a pipe).

Multiple WUs can also join their efforts to pursue common interventions,
for example, the cost of placing a pipe connecting two water utilities
will be shared between the two utilities. If WU are connected with each
other and there is an exchange of water, the receiving utility pays a
volumetric fee to the supplying utility. More detailed information about
the water tariff scheme are available in
Section~\ref{sec:water-pricing}.

The structure of the system is static. The number of WUs and their
specific geographic responsibilities (assigned provinces) will not
change through the competition.

The key parameters and decision variables governing the Water Utility
Module are detailed in Table~\ref{tbl:wu-properties}. The actual values
for these variables can be inspected within the data files, which are
mapped in Appendix A.

\hypertarget{tbl:wu-properties}{}
\begin{longtable}[]{@{}llll@{}}
\caption{\label{tbl:wu-properties}Water utilities' properties
review.}\tabularnewline
\toprule\noalign{}
Property & Type & Scope & Unit \\
\midrule\noalign{}
\endfirsthead
\toprule\noalign{}
Property & Type & Scope & Unit \\
\midrule\noalign{}
\endhead
\bottomrule\noalign{}
\endlastfoot
Identifier & Static & Water Utility & \\
Assigned province & Static & Water Utility & \\
\end{longtable}

\hypertarget{municipalities}{%
\subsubsection{Municipalities}\label{municipalities}}

Each municipality (gemeente) is represented as a single junction point
with positive demand, abstracting the entire secondary and tertiary
distribution network within that jurisdiction. This node serves as the
sole supply point for all water within the municipality's boundaries,
with characteristics such as population, land area, and housing stock
consolidated at this level.

The system presents two main challenges. First, municipal parameters
evolve over time as cities grow and change. Second, the network topology
itself is dynamic: municipalities can merge or be absorbed by larger
neighbors, causing the number of nodes to vary throughout the planning
horizon.

\begin{center}\rule{0.5\linewidth}{0.5pt}\end{center}

\textbf{Excursus on the Modelling Approach}

\begin{figure}
\hypertarget{fig:municipality-dissolution}{%
\centering
\includegraphics[width=0.8\textwidth,height=\textheight]{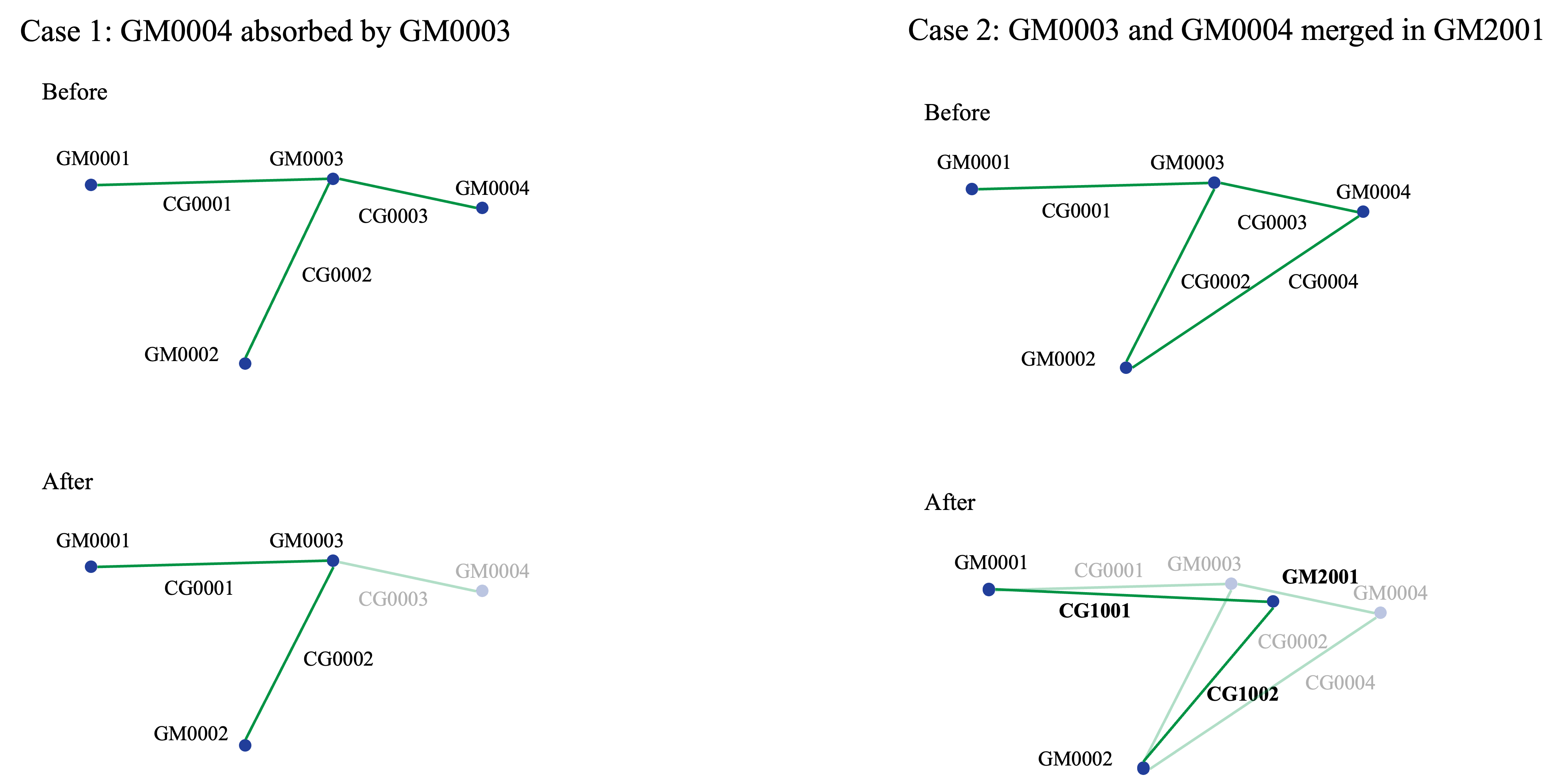}
\caption{Network topology before and after municipal dissolution
(dissolved entities are greyed out). Case 1 shows absorption of GM0004
by the existing municipality GM0003: the inter-municipal connection
CG0003 becomes a self-loop and is hidden. Case 2 illustrates the
clustering of GM0003 and GM0004 into the new municipality GM2001, where
CG1001 replaces CG0001 as a one-to-one substitution, while CG1002
consolidates the previously separate connections CG0002 and CG0004 into
a single new link.}\label{fig:municipality-dissolution}
}
\end{figure}

To model this administrative restructuring, municipalities can only open
or close on January 1st of each year. When a municipality closes, its
delivery point disappears from the network and its assets (population,
land area, housing stock, etc.) are redistributed between the
``destination municipalities'', with the ``main destination
municipality'' inheriting also the hydraulic connections to other
municipalities and water sources. Extensive properties (e.g.,
population, number of houses) are accumulated, while intensive
properties (e.g., average age of the inner distribution network) are
distributed using a weighted mean. Besides renaming,
Figure~\ref{fig:municipality-dissolution} illustrates the two possible
cases for dissolved municipalities\footnote{Dissolved municipalities are
  referred to as `lifted' (`opgeheven') in the input data.}, which are:

\begin{enumerate}
\def\labelenumi{\arabic{enumi}.}
\item
  \emph{Absorption by existing municipalities}: When a municipality is
  absorbed by a larger neighbor that already exists, all attributes of
  the closing municipality transfer to the destination municipality. Any
  pipe that previously connected these two entities becomes hidden, as
  it formally becomes part of the destination municipality's internal
  distribution network\footnote{These connections have the
    ``SELF\_LOOP'' label in the ``replaced\_by'' column in the input
    data.}.
\item
  \emph{Clustering into new municipalities}: When multiple
  municipalities close and cluster together to form a new entity, all
  their delivery points disappear and a new supply point emerges at the
  location of the newly formed municipality. Of course, the new
  municipality attributes are computed by aggregating those of its
  constituent municipalities. Internal connections between merging
  municipalities become hidden as for the ``Absorption by existing
  municipalities'' case. External connections to neighbour
  municipalities are replaced by new connections routed to the new city
  centre. While it's not possible to operate on the original connections
  anymore, the old connection remains active as a fallback to preserve
  network connectivity.
\end{enumerate}

This modelling approach mirrors real-world dynamics in densely populated
countries like the Netherlands. For example, when a new municipality
forms through clustering, typically a new city center is established
while former city centers become secondary neighborhoods. These moments
of urban reorganization present natural opportunities for water
utilities to lay new connections and redesign substantial portions of
the distribution system.

\begin{center}\rule{0.5\linewidth}{0.5pt}\end{center}

Municipalities have many attributes that influence the other modules of
the system. The full list can be seen in
Table~\ref{tbl:muni-properties}, while the actual values for these
variables can be inspected within the data files, which are mapped in
Appendix A.

\hypertarget{tbl:muni-properties}{}
\begin{longtable}[]{@{}
  >{\raggedright\arraybackslash}p{(\columnwidth - 6\tabcolsep) * \real{0.2500}}
  >{\raggedright\arraybackslash}p{(\columnwidth - 6\tabcolsep) * \real{0.2500}}
  >{\raggedright\arraybackslash}p{(\columnwidth - 6\tabcolsep) * \real{0.2500}}
  >{\raggedright\arraybackslash}p{(\columnwidth - 6\tabcolsep) * \real{0.2500}}@{}}
\caption{\label{tbl:muni-properties}Municipalities' properties
review.}\tabularnewline
\toprule\noalign{}
\begin{minipage}[b]{\linewidth}\raggedright
Property
\end{minipage} & \begin{minipage}[b]{\linewidth}\raggedright
Type
\end{minipage} & \begin{minipage}[b]{\linewidth}\raggedright
Scope
\end{minipage} & \begin{minipage}[b]{\linewidth}\raggedright
Unit
\end{minipage} \\
\midrule\noalign{}
\endfirsthead
\toprule\noalign{}
\begin{minipage}[b]{\linewidth}\raggedright
Property
\end{minipage} & \begin{minipage}[b]{\linewidth}\raggedright
Type
\end{minipage} & \begin{minipage}[b]{\linewidth}\raggedright
Scope
\end{minipage} & \begin{minipage}[b]{\linewidth}\raggedright
Unit
\end{minipage} \\
\midrule\noalign{}
\endhead
\bottomrule\noalign{}
\endlastfoot
Name & Static & Municipality & \\
Identifier & Static & Municipality & \\
Latitude & Static & Municipality & degrees \\
Longitude & Static & Municipality & degrees \\
Elevation & Static & Municipality & m \\
Province & Static & Municipality & \\
Begin date & Static & Municipality & date \\
End date & Static {[}Optional{]} & Municipality & date \\
End reason and destination & Static {[}Optional{]} & Municipality & \\
Population & Dynamic Exogenous & Municipality & inhabitants \\
Surface land & Dynamic Exogenous & Municipality & \(km^2\) \\
Surface water (inland) & Dynamic Exogenous & Municipality & \(km^2\) \\
Surface water (open water) & Dynamic Exogenous & Municipality &
\(km^2\) \\
Number of houses & Dynamic Exogenous & Municipality & units \\
Number of businesses & Dynamic Exogenous & Municipality & units \\
Average Disposable Income & Dynamic Exogenous & Municipality &
\(k\text{€}\) \\
Average age distribution network & Dynamic Endogenous & Municipality &
years \\
\end{longtable}

The total municipality water demand comprises two volumetric quantities:

\begin{itemize}
\tightlist
\item
  Billable water demand (\(D^\text{BIL}\)): The sum of household and
  business water demands described in Section~\ref{sec:water-dem}.
\item
  Non-revenue water (\(D^\text{NRW}\)): accounting for leaks, flushing,
  measurement errors and other losses described in
  Section~\ref{sec:nrw}.
\end{itemize}

While these quantities represent the physical components of demand, they
are not directly observable by participants. Instead, participants
observe the total water demand divided into two components: consumption
(\(Q\); delivered outflow) and undelivered demand (\(U\)).

For each municipality \(m\) at time \(t\), these quantities maintain the
following relationships:

\begin{equation}\protect\hypertarget{eq:comp-of-demands}{}{
\begin{aligned}
D_m(t) &= D^\text{BIL}_m(t) + D^\text{NRW}_m(t) \\
 &= Q_m(t) + U_m(t)
\end{aligned}
}\label{eq:comp-of-demands}\end{equation}

The decomposition between delivered and undelivered demand is extracted
from an EPANET simulation of the network run in pressure-driven analysis
(PDA) mode with a minimum pressure threshold of 30 m. Whenever there is
undelivered demand, we assume that this reduces the billable component
first, i.e., \(Q^\text{BIL}_m(t) = D^\text{BIL}_m(t) - U_m(t)\).

\hypertarget{sec:water-dem}{%
\paragraph{Water Demand Model}\label{sec:water-dem}}

The methodology developed to generate water consumption time series
builds on historical data from the Dutch association of water companies
(Vewin 2025), which provide nationwide trends in total drinking water
production, sectoral water use, and non-revenue water over the period
2000--2024. Specifically, water-consumption time series generation is
structured into three phases.

Phase I. The first phase estimates the annual water volume supplied to
each municipality using information on households and businesses (CBS
2025), complemented by projected data where required. These annual
volumes are calibrated to match national totals reported in official
statistics (Vewin 2025) and then randomized around the calibrated value
to introduce variability among municipalities.

Phase II. In the second phase, representative hourly consumption
profiles are assigned to each municipality using a library of year-long,
normalized profiles derived from district-metered areas and
pre-processed to remove leakage effects. In greater detail, for each
municipality, two residential profiles are selected from the library
according to municipality population class, while a single
non-residential profile is drawn from a dedicated set.

Phase III. The third phase produces the final hourly time series by
applying a Fourier series-based approach which combines seasonal
modulation, climate-related adjustments (accounting for the maximum
yearly temperature), and random perturbations to capture temporal
variability. The two residential profiles associated with each
municipality are aggregated through weighted combinations (with the
weights being uncertain), and both residential and non-residential
profiles are scaled to match the previously estimated yearly volumes.

Therefore, the total billable demand of municipality \(m\) at time \(t\)
(within year \(y\)) is defined as:

\begin{equation}\protect\hypertarget{eq:water-demand-model}{}{
D^\text{BIL}_m(t) = D^\text{R1}_m(t, T_y) \cdot w_m + D^\text{R2}_m(t, T_y) \cdot (1-w_m) + D^\text{C}_m(t, T_y)
}\label{eq:water-demand-model}\end{equation}

where \(D^\text{R1}_m(t, T_y)\) and \(D^\text{R2}_m(t, T_y)\) represent
the two residential demands, \(w_m \in [0,1]\) is the unitary weight to
combine them, \(D^\text{C}_m(t, T_y)\) is the (commercial)
non-residential demand, and \(T_y\) is the maximum temperature recorded
in year \(y\).

\hypertarget{tbl:wd-properties}{}
\begin{longtable}[]{@{}
  >{\raggedright\arraybackslash}p{(\columnwidth - 6\tabcolsep) * \real{0.2500}}
  >{\raggedright\arraybackslash}p{(\columnwidth - 6\tabcolsep) * \real{0.2500}}
  >{\raggedright\arraybackslash}p{(\columnwidth - 6\tabcolsep) * \real{0.2500}}
  >{\raggedright\arraybackslash}p{(\columnwidth - 6\tabcolsep) * \real{0.2500}}@{}}
\caption{\label{tbl:wd-properties}Water demand model's properties
review.}\tabularnewline
\toprule\noalign{}
\begin{minipage}[b]{\linewidth}\raggedright
Property
\end{minipage} & \begin{minipage}[b]{\linewidth}\raggedright
Type
\end{minipage} & \begin{minipage}[b]{\linewidth}\raggedright
Scope
\end{minipage} & \begin{minipage}[b]{\linewidth}\raggedright
Unit
\end{minipage} \\
\midrule\noalign{}
\endfirsthead
\toprule\noalign{}
\begin{minipage}[b]{\linewidth}\raggedright
Property
\end{minipage} & \begin{minipage}[b]{\linewidth}\raggedright
Type
\end{minipage} & \begin{minipage}[b]{\linewidth}\raggedright
Scope
\end{minipage} & \begin{minipage}[b]{\linewidth}\raggedright
Unit
\end{minipage} \\
\midrule\noalign{}
\endhead
\bottomrule\noalign{}
\endlastfoot
Population & Dynamic Exogenous & Municipality & inhabitants \\
Number of houses & Dynamic Exogenous & Municipality & units \\
Number of businesses & Dynamic Exogenous & Municipality & units \\
Daily per household demand & Dynamic Exogenous & Municipality &
m³/house/hour \\
Daily per business demand & Dynamic Exogenous & Municipality &
m³/business/hour \\
Max yearly temperature & Dynamic Exogenous & National & °C \\
\end{longtable}

\hypertarget{sec:nrw}{%
\paragraph{Non-Revenue Water Model}\label{sec:nrw}}

Non-revenue water (NRW) is an uncertain quantity modeled through the
average age of pipe infrastructure in each municipality's inner
distribution network (IDN). Based on this average age, municipalities
are assigned to one of five NRW classes as reported in
Table~\ref{tbl:nrw-classes}. Each class is associated with a distinct
probability distribution of NRW demands, from which daily samples are
drawn to generate the volumetric NRW demand factor (\(m^3/km/day\)).
Notably, older infrastructure suffers from more leaks and therefore
exhibits higher NRW demand factors.

\begin{figure}
\hypertarget{fig:nrw-demand}{%
\centering
\includegraphics[width=0.6\textwidth,height=\textheight]{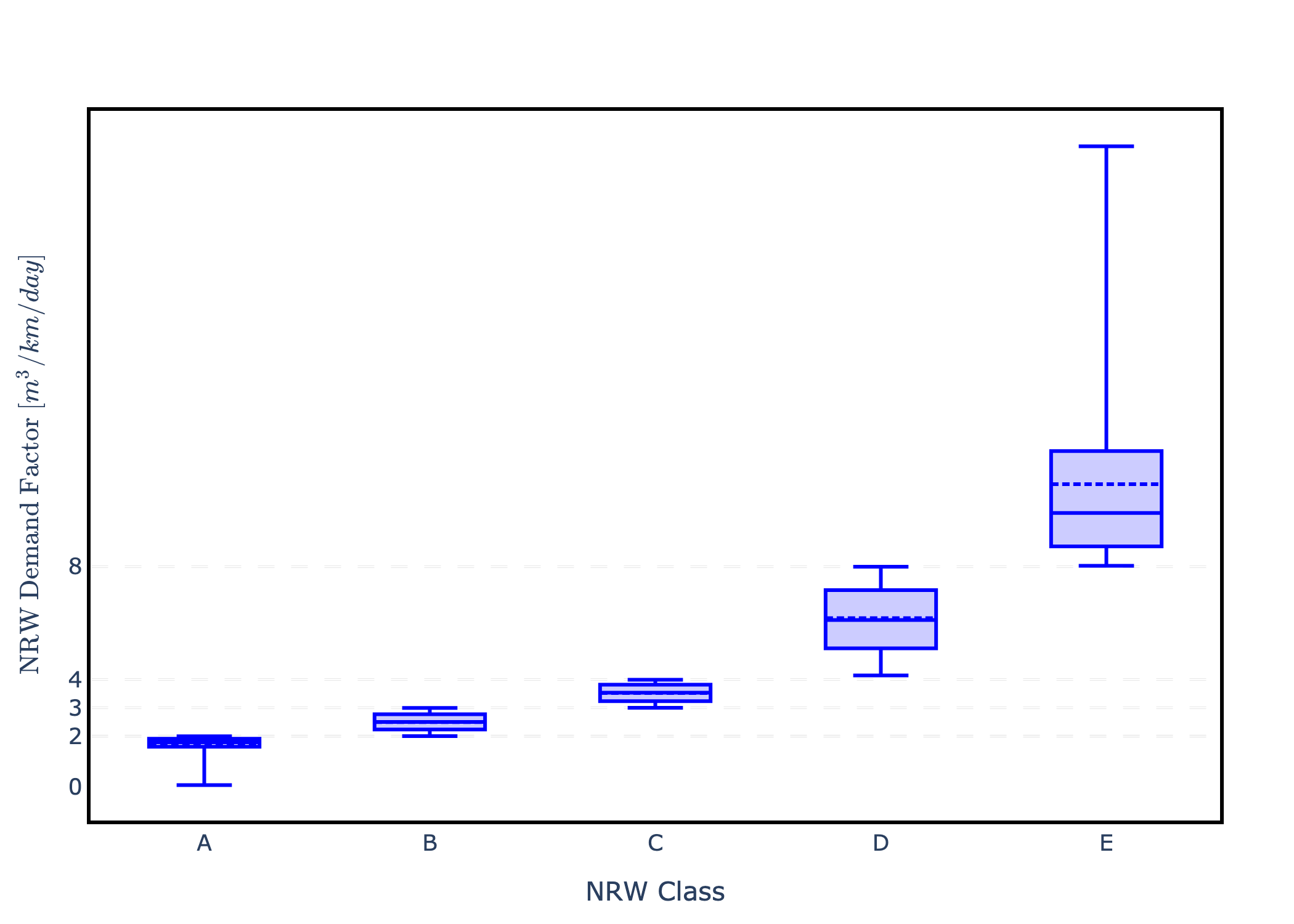}
\caption{Non-revenue water demand factor per
class}\label{fig:nrw-demand}
}
\end{figure}

The distribution of NRW demands varies by class and is illustrated in
Figure~\ref{fig:nrw-demand}. The actual values of the distributions'
parameters can be inspected within the data files, which are mapped in
Appendix A.

The total length of pipes in a municipality is linked to its population
size through the following linear relationship:

\begin{equation}\protect\hypertarget{eq:pipe-length}{}{
L^\text{IDN}_{m}(y) = 57.7*10^{-4} \cdot \text{inhabitants}_{m}(y) 
}\label{eq:pipe-length}\end{equation}

where \(m\) is the municipality index, \(y\) is the year, and
\(L^\text{IDN}_{m}(y)\) is the total length of pipes (km) in
municipality \(m\) at year \(y\).

The actual municipality NRW demand is also capped at twice the billable
daily demand to prevent unrealistic leakage levels. Therefore, total NRW
demand for municipality \(m\) at day \(d\) is calculated as:

\begin{equation}\protect\hypertarget{eq:new-demand}{}{
D^\text{NRW}_{m}(d) = \min\left(f^\text{NRW}_{\text{class}(m)} \cdot L^\text{IDN}_{m}(y), \, 2 \cdot \bar D^\text{BIL}_{m}(d)\right)
}\label{eq:new-demand}\end{equation}

where \(f^\text{NRW}_{\text{class}(m)}\)\hspace{0pt} is the sampled NRW
demand factor (\(m^3/km/day\)) for the municipality's class, and
\(\bar D^\text{BIL}_{m}(d)\)\hspace{0pt} is the average daily water
demand of municipality \(m\) at day \(d\). The daily leak
\(D^\text{NRW}_{m}(d)\) is equally spread across the day (i.e.,
\(D^\text{NRW}_{m}(t)=D^\text{NRW}_{m}(d)/24\)).

\hypertarget{tbl:nrw-classes}{}
\begin{longtable}[]{@{}
  >{\raggedright\arraybackslash}p{(\columnwidth - 4\tabcolsep) * \real{0.3000}}
  >{\raggedright\arraybackslash}p{(\columnwidth - 4\tabcolsep) * \real{0.3000}}
  >{\raggedright\arraybackslash}p{(\columnwidth - 4\tabcolsep) * \real{0.4000}}@{}}
\caption{\label{tbl:nrw-classes}Non-revenue water classification by
infrastructure age.}\tabularnewline
\toprule\noalign{}
\begin{minipage}[b]{\linewidth}\raggedright
Inner distribution network - average age {[}years{]}
\end{minipage} & \begin{minipage}[b]{\linewidth}\raggedright
NRW Class
\end{minipage} & \begin{minipage}[b]{\linewidth}\raggedright
Probability distribution
\end{minipage} \\
\midrule\noalign{}
\endfirsthead
\toprule\noalign{}
\begin{minipage}[b]{\linewidth}\raggedright
Inner distribution network - average age {[}years{]}
\end{minipage} & \begin{minipage}[b]{\linewidth}\raggedright
NRW Class
\end{minipage} & \begin{minipage}[b]{\linewidth}\raggedright
Probability distribution
\end{minipage} \\
\midrule\noalign{}
\endhead
\bottomrule\noalign{}
\endlastfoot
0 - 25 & A & Inverted Exponential \\
25 - 43 & B & Uniform \\
43 - 54 & C & Uniform \\
54 - 60 & D & Uniform \\
\textgreater{} 60 & E & Exponential \\
\end{longtable}

\hypertarget{tbl:nrw-properties}{}
\begin{longtable}[]{@{}
  >{\raggedright\arraybackslash}p{(\columnwidth - 6\tabcolsep) * \real{0.2500}}
  >{\raggedright\arraybackslash}p{(\columnwidth - 6\tabcolsep) * \real{0.2500}}
  >{\raggedright\arraybackslash}p{(\columnwidth - 6\tabcolsep) * \real{0.2500}}
  >{\raggedright\arraybackslash}p{(\columnwidth - 6\tabcolsep) * \real{0.2500}}@{}}
\caption{\label{tbl:nrw-properties}Non-revenue water model's properties
review.}\tabularnewline
\toprule\noalign{}
\begin{minipage}[b]{\linewidth}\raggedright
Property
\end{minipage} & \begin{minipage}[b]{\linewidth}\raggedright
Type
\end{minipage} & \begin{minipage}[b]{\linewidth}\raggedright
Scope
\end{minipage} & \begin{minipage}[b]{\linewidth}\raggedright
Unit
\end{minipage} \\
\midrule\noalign{}
\endfirsthead
\toprule\noalign{}
\begin{minipage}[b]{\linewidth}\raggedright
Property
\end{minipage} & \begin{minipage}[b]{\linewidth}\raggedright
Type
\end{minipage} & \begin{minipage}[b]{\linewidth}\raggedright
Scope
\end{minipage} & \begin{minipage}[b]{\linewidth}\raggedright
Unit
\end{minipage} \\
\midrule\noalign{}
\endhead
\bottomrule\noalign{}
\endlastfoot
Inner distribution network - length to population ratio & Static &
National & \(km \cdot (10^4 \text{ inhabitants})^{-1}\) \\
Inner distribution network - length & Dynamic endogenous & Municipality
& \(km\) \\
Inner distribution network - average age & Dynamic endogenous &
Municipality & \(years\) \\
NRW intervention - unit cost & Dynamic endogenous & NRWClass,
Municipality Size Class, National & \(\text{€}/year/km\) \\
NRW intervention - effectiveness factor & Static {[}Uncertain{]} &
NRWClass, Municipality Size Class, National & \\
Intervention budget & Option & & \(\text{€}/year\) \\
Intervention policy & Option & & \\
\end{longtable}

\hypertarget{sec:water-sources}{%
\subsubsection{Water Sources}\label{sec:water-sources}}

Water sources represent the entire process of water abstraction and
treatment consolidated at a single location. Each source has a nominal
production capacity which represents a hard limit on the amount of water
that can be delivered throughout any consecutive 24-hour window.

In the BWF, three macro-types of water sources are modelled:
groundwater, surface water, and desalination. The main trade-off is that
groundwater sources are generally smaller (\textless{} 25 \(Mm^3/year\))
and cheaper to operate, as the water has higher quality and requires
less treatment. Surface water treatment plants and desalination plants
are generally much larger (30-45 \(Mm^3/year\)) but are considerably
more expensive and energy-intensive to run. Moreover, surface water
sources differ as they are affected by climate conditions; specifically,
low inflows within rivers could temporarily shut down treatment plants
downstream (effectively reducing their capacity to zero for those days).
Instead, groundwater sources have an extraction permit (expressed in
\(m^3\) per year). This is not a hard physical constraint such as the
nominal production capacity, but rather a ``soft'' legislative
constraint checked by the government at the end of each year. This
penalty effectively represents compensation for the hydrological
displacement affecting farmers and natural areas due to the
overextraction. The fine amount is set by law and can therefore change
at any time based on political decisions. The displacement thresholds
for the application of the fine are presented in
Table~\ref{tbl:gwsources-permit-violation}.

The cost of water production at every source \(s\) is a combination of
four components:

\begin{equation}\protect\hypertarget{eq:water-production-cost}{}{
\mathrm{OPEX}_s(y) =  F_s + c^{\text{el}}(y) \cdot E^{\text{grid}}_s(y) +c^\text{vol}_s(y) \cdot Q_s(y) + c^\text{extra}_s(y) \cdot \max\bigl(Q_s(y)-\phi_s \cdot Q^*_s, 0 \bigr)
}\label{eq:water-production-cost}\end{equation}

where:

\begin{itemize}
\tightlist
\item
  \(F_s\) represents fixed costs, including personnel, taxes, and
  planned maintenance.
\item
  \(c^\text{el}(y)\cdot E^{\text{grid}}_s(y)\) represents volumetric
  costs for energy.
\item
  \(c^\text{vol}_s(y) \cdot Q_s(y)\) represents the volumetric costs for
  non-energy related expenditures, such as chemicals and filters.
\item
  \(c^\text{extra}_s(y) \cdot \max\bigl(Q_s(y)-\phi_s \cdot Q^*_s, 0 \bigr)\)
  represent the extra volumetric costs incurred when production exceeds
  the planned threshold.
\end{itemize}

In Equation~\ref{eq:water-production-cost}, \(Q_s(y)\) is the total
volume produced by the source in year \(y\), \(Q^*_s\) is the source
nominal capacity, and \(\phi_s\) is the capacity target factor
(source-type dependent), which defines the ideal operating point above
which additional costs are applied.

Additionally, the variable \(E^{\text{grid}}_s(y)\) represents the
electrical grid energy consumed for water production by the source \(s\)
in year \(y\) and it is calculated as:
\begin{equation}\protect\hypertarget{eq:source-energy-grid}{}{
E^{\text{grid}}_s(y) = \sum_{t \in \mathcal{Y}} \max\bigl(E_s(t) - E^{\text{PV}}_s(t), 0\bigr)
}\label{eq:source-energy-grid}\end{equation}

where \(t\) is the simulation timestep, \(\mathcal{Y}\) is the set of
timesteps in year \(y\), \(E_s(t)\) represents the energy consumed by
the sources \(s\) at time \(t\), and the \(E^{\text{PV}}_s(t)\) is
electricity generated by the photovoltaic panels associated with the
source \(s\) at time \(t\).

The source energy consumption \(E_s(t)\) is:

\begin{equation}\protect\hypertarget{eq:source-energy}{}{
E_s(t) = \epsilon_s \cdot Q_s(t)
}\label{eq:source-energy}\end{equation}

where \(\epsilon_s\) is the source specific energy and \(Q_s(t)\) is the
volume produced by the source at timestep \(t\).

To complete the accounting, the total operational expenditure associated
with all the sources in water utility \(w\) at year \(y\) is:

\begin{equation}\protect\hypertarget{eq:op-expends}{}{
\mathrm{OPEX}^{\text{sources}}_w(y) = \sum_{s \in \mathcal{S}_w} \mathrm{OPEX}_s(y)
}\label{eq:op-expends}\end{equation}

where \(\mathcal{S}_w\) represents the collection of water sources
managed by utility \(w\), and \(\mathrm{OPEX}_s(y)\) is the individual
source cost defined in Equation~\ref{eq:water-production-cost}.

Similarly to municipalities, sources can also open and close over time.
Participants can decide to close production locations and open new ones
within the constraints of the problem (available locations and sizes) to
make the supply system more efficient. However, a closed source cannot
be reopened and no direct cost is associated with this action. When
activating a new source, participants must decide the nominal capacity,
but the possible size is limited by different rules depending on the
source type:

\begin{itemize}
\tightlist
\item
  Desalination and surface water sources have an upper limit defined in
  the competition data (see Appendix A)
\item
  Groundwater sources cannot exceed the permit by more than 30\%
\end{itemize}

New water sources have an uncertain construction time, so participants
must communicate the construction start date, and the activation date
will be randomized.

The capital investment associated with the construction of new sources
in water utility \(w\) at year \(y\) is:

\begin{equation}\protect\hypertarget{eq:capex-sources}{}{
\text{CAPEX}^\text{sources}_w(y)
= \sum_{s \in \mathcal{S}_w} \mathbf{1}_{\{\tau_s = y\}} \cdot c^\text{source} (\text{class}(s),y) \cdot Q^*_s 
}\label{eq:capex-sources}\end{equation}

where for a source \(s\) in the set of the water utility's sources
(\(\mathcal{S}_w\)), \(\tau_s\) is its starting construction time,
\(\mathbf{1}_{\{\tau_s = y\}}\) is an indicator function equal to 1 if
the construction begins in year \(y\) (0 otherwise), \(Q^*_s\) is the
requested source nominal capacity, and
\(c^\text{source} (\text{class}(s),y)\) the unit cost, which depends on
the year of construction and the source class (see
Table~\ref{tbl:sources-classes}).

\hypertarget{tbl:sources-classes}{}
\begin{longtable}[]{@{}ll@{}}
\caption{\label{tbl:sources-classes}Sources classification by their
nominal capacity.}\tabularnewline
\toprule\noalign{}
Source nominal capacity {[}\(Mm^3/year\){]} & Source Class \\
\midrule\noalign{}
\endfirsthead
\toprule\noalign{}
Source nominal capacity {[}\(Mm^3/year\){]} & Source Class \\
\midrule\noalign{}
\endhead
\bottomrule\noalign{}
\endlastfoot
4 & SMALL \\
8 & MEDIUM \\
16 & LARGE \\
\textgreater{} 16 & VERY LARGE \\
\end{longtable}

One limitation of the BWF is that we do not model water quality
differences between sources, which would typically prevent mixing in
practice. This simplification keeps the problem tractable given its
existing complexity.

The key parameters and decision variables governing the Water Sources
Module are detailed in Table~\ref{tbl:sources-properties}. The actual
values for these variables can be inspected within the data files, which
are mapped in Appendix A.

\hypertarget{tbl:sources-properties}{}
\begin{longtable}[]{@{}
  >{\raggedright\arraybackslash}p{(\columnwidth - 6\tabcolsep) * \real{0.2500}}
  >{\raggedright\arraybackslash}p{(\columnwidth - 6\tabcolsep) * \real{0.2500}}
  >{\raggedright\arraybackslash}p{(\columnwidth - 6\tabcolsep) * \real{0.2500}}
  >{\raggedright\arraybackslash}p{(\columnwidth - 6\tabcolsep) * \real{0.2500}}@{}}
\caption{\label{tbl:sources-properties}Sources' properties
review.}\tabularnewline
\toprule\noalign{}
\begin{minipage}[b]{\linewidth}\raggedright
Property
\end{minipage} & \begin{minipage}[b]{\linewidth}\raggedright
Type
\end{minipage} & \begin{minipage}[b]{\linewidth}\raggedright
Scope
\end{minipage} & \begin{minipage}[b]{\linewidth}\raggedright
Unit
\end{minipage} \\
\midrule\noalign{}
\endfirsthead
\toprule\noalign{}
\begin{minipage}[b]{\linewidth}\raggedright
Property
\end{minipage} & \begin{minipage}[b]{\linewidth}\raggedright
Type
\end{minipage} & \begin{minipage}[b]{\linewidth}\raggedright
Scope
\end{minipage} & \begin{minipage}[b]{\linewidth}\raggedright
Unit
\end{minipage} \\
\midrule\noalign{}
\endhead
\bottomrule\noalign{}
\endlastfoot
Name & Static {[}Optional{]} & Source & \\
Identifier & Static & Source & \\
Source type & Static & Source & \\
Latitude & Static & Source & degrees \\
Longitude & Static & Source & degrees \\
Elevation & Static & Source & m \\
Province & Static & Source & \\
Connected municipality & Static & Source & \\
Activation date & Static & Source & date \\
Closure date & Static {[}Optional{]} {[}Decision{]} & Source & date \\
Capacity - nominal & Static {[}Optional{]} {[}Decision{]} & Source &
\(m^3/day\) \\
Capacity - target factor & Static & Source Type & \% \\
Unit cost of construction & Dynamic Endogenous & Source Size Class
\(\times\) National & \(\text{€}/(m^3/day)\) \\
Operational costs - fixed & Dynamic Endogenous {[}Uncertain{]} & Source
Size Class \(\times\) National & \(\text{€}/year\) \\
Specific energy & Static {[}Uncertain{]} & Source & \(kWh/m^3\) \\
Operational costs - volumetric for non-energy & Dynamic Endogenous
{[}Uncertain{]} & Source Size Class \(\times\) National &
\(\text{€}/m^3\) \\
Operational costs - volumetric for non-energy - multiplier & Static &
Source Type & \% \\
Construction time & Static {[}Uncertain{]} & Source Type & years \\
Availability factor & Dynamic Exogenous & Surface water sources & \\
Permit & Static & Groundwater sources & \(m^3/year\) \\
Fine amount & Dynamic Exogenous & Groundwater source permit violation
Severity Class & \(\text{€}\) \\
\end{longtable}

\hypertarget{tbl:gwsources-permit-violation}{}
\begin{longtable}[]{@{}ll@{}}
\caption{\label{tbl:gwsources-permit-violation}Groundwater sources'
permit violation severity classes.}\tabularnewline
\toprule\noalign{}
Permit violation {[}\(Mm^3/year\){]} & Severity Class \\
\midrule\noalign{}
\endfirsthead
\toprule\noalign{}
Permit violation {[}\(Mm^3/year\){]} & Severity Class \\
\midrule\noalign{}
\endhead
\bottomrule\noalign{}
\endlastfoot
0.0 & COMPLIANT \\
\textless{} 0.1 & MILD \\
\textless{} 8.0 & SEVERE \\
\textgreater{} 8.0 & EXTREME \\
\end{longtable}

\hypertarget{pumping-stations}{%
\subsubsection{Pumping stations}\label{pumping-stations}}

Pumping stations are the infrastructure connecting water sources to
municipalities and are responsible for distributing the water produced
by each source into the network. Each pumping station contains one or
more identical pumps operating in parallel.

Whenever a new source is opened, participants must define which pump
option (selected exclusively from a pre-defined list of available pumps)
and how many units are installed at that source's pumping station.
Similarly, participants can decide to replace pumps at existing
locations or add new ones of the same type as installed.

Whether replacing old pumps or being added in new pumping stations, the
capital investment associated with the installation of pumps in water
utility \(w\) at year \(y\) is:

\begin{equation}\protect\hypertarget{eq:pumping-stations-capital-cost}{}{
\text{CAPEX}^\text{pumps}_w(y)
= \sum_{n \in \mathcal{N}_w} \sum_{p \in \mathcal{P}_{n}} \mathbf{1}_{\{\tau_p = y\}}\cdot c^\text{pump}(\gamma_p, y)
}\label{eq:pumping-stations-capital-cost}\end{equation}

where \(n\) is the pumping station index within the set of the water
utility's pumping stations (\(\mathcal{N}_w\)), \(p\) is the p-th pump
in that pumping station set (\(\mathcal{P}_n\)), \(\tau_p\) is the pump
installation time, \(\mathbf{1}_{\{\tau_p = y\}}\) is an indicator
function equal to 1 if the installation happened in year \(y\) (0
otherwise), and \(c^\text{pump}(\gamma_p, y)\) the unit cost for the
selected pump option \(\gamma_p\).

While a water source's daily outflow is constrained by its nominal
capacity, the pumping station's configuration limits the source's peak
outflow rate (\(m^3/hour\)).

The only operational cost component of a pumping station is its energy
expenditure, calculated during the network hydraulic simulation based on
the energy consumption and tariffs. Peak demand charges are not
included, as we assume utilities have agreements in place with
electrical grid providers (see more details in
Section~\ref{sec:energy-model}). Maintenance costs and other fixed
yearly operational costs are included in the initial construction cost
of each pump.

Thus, the operational expenditure associated with pumping in water
utility \(w\) at year \(y\) is:

\begin{equation}\protect\hypertarget{eq:pumping-stations-opex}{}{
\mathrm{OPEX}^{\text{pumps}}_w(y) = \sum_{n \in \mathcal{N}_w} c^{\text{el}}(y) \cdot E^{\text{grid}}_n(y)
}\label{eq:pumping-stations-opex}\end{equation}

where \(n\) is the pumping station index within the set of the water
utility's pumping stations (\(\mathcal{N}_w\)), \(c^\text{el}(y)\) is
the dynamic electricity price, and \(E^{\text{grid}}_n(y)\) is the
electricity consumption from the grid for the \emph{n}-th pumping
station in year \(y\). This quantity is computed as:

\begin{equation}\protect\hypertarget{eq:pumping-stations-energy}{}{
E^{\text{grid}}_n(y) = \sum_{t \in \mathcal{Y}} \max\bigl( \sum_{p \in \mathcal{P}_{n}} E_p(t) - E^{\text{PV}}_n(t), 0\bigr)
}\label{eq:pumping-stations-energy}\end{equation}

where \(t\) is the simulation timestep, \(\mathcal{Y}\) is the set of
timesteps in year \(y\), \(E_p(t)\) the \emph{p}-th pump energy
consumption at time \(t\), \(\mathcal{P}_n\) the set of pumps in pumping
station \(n\), and \(E^\text{PV}_n(t)\) is electricity generated by the
photovoltaic panels associated with pumping station \(n\) at time \(t\).

Pumps performances remain constant over time (no degradation in
efficiency or similar wear effects). However, pumps do age normally and
have an expected lifetime. The actual lifetime is randomized, and when a
pump reaches its end of life, it must be replaced. Participants do not
need to communicate this decision, as replacements will be automatically
implemented, but they must account for this ``unexpected'' cost in their
planning.

\hypertarget{tbl:ps-properties}{}
\begin{longtable}[]{@{}llll@{}}
\caption{\label{tbl:ps-properties}Pumping stations' properties
review.}\tabularnewline
\toprule\noalign{}
Property & Type & Scope & Unit \\
\midrule\noalign{}
\endfirsthead
\toprule\noalign{}
Property & Type & Scope & Unit \\
\midrule\noalign{}
\endhead
\bottomrule\noalign{}
\endlastfoot
Identifier & Static & Pumping station & \\
Assigned source & Static & Pumping station & \\
Pump option & Decision & Pumping station & \\
Number of pumps & Decision & Pumping station & unit \\
\end{longtable}

\hypertarget{tbl:pu-properties}{}
\begin{longtable}[]{@{}llll@{}}
\caption{\label{tbl:pu-properties}Pumps' properties
review.}\tabularnewline
\toprule\noalign{}
Property & Type & Scope & Unit \\
\midrule\noalign{}
\endfirsthead
\toprule\noalign{}
Property & Type & Scope & Unit \\
\midrule\noalign{}
\endhead
\bottomrule\noalign{}
\endlastfoot
Identifier & Static & Pump option & \\
Pump curve & Static & Pump option & \\
Efficiency curve & Static & Pump option & \\
Lifetime & Static {[}Uncertain{]} & Pump option & years \\
Unit cost & Dynamic Endogenous & Pump option & € \\
\end{longtable}

Note that the provided pump curves only describe the working range of
the pumps and must not be extrapolated beyond the values presented in
the tables.

\hypertarget{sec:connections}{%
\subsubsection{Connections (Pipes)}\label{sec:connections}}

Within the BWF, there is a distinction between connections and pipes.
Connections are the entities that define the possible links between the
network's nodes, while pipes are the actual physical elements installed
to transport water. This distinction models the fact that different
pipes can be installed on the same connection at different points in
time. However, duplicate pipes are not allowed on the same connection.

Similarly to pumps, participants can install new pipes on any unused
connection, or replace existing pipes on connections that already have
one. Selection must occur from a predefined set of pipe options. Pipe
options are characterized by material, hydraulic properties,
installation cost, and associated carbon emissions, with a detailed
overview available in Table~\ref{tbl:pi-properties} and actual values
provided in the data files mapped in Appendix A.

Whether replacing old pipes or being added in a connection, the capital
investment associated with the installation of pipes in water utility
\(w\) at year \(y\) is:

\begin{equation}\protect\hypertarget{eq:pipes-capital-cost}{}{
\text{CAPEX}^\text{pipes}_w(y)
= \sum_{j \in \mathcal{J}_{w}} \mathbf{1}_{\{\tau_j = y\}}\cdot c^\text{pipe}(\theta_j, y) \cdot L_j
}\label{eq:pipes-capital-cost}\end{equation}

where \(\mathcal{J}_{w}\) is the set of all pipes within water utility
\(w\), \(j\) the pipe index, \(\tau_j\) is the pipe installation time,
\(\mathbf{1}_{\{\tau_j = y\}}\) is an indicator function equal to 1 if
the installation happened in year \(y\) (0 otherwise),
\(c^\text{pipe}(\theta_j, y)\) the unit cost for the selected pipe
option \(\theta_j\), and \(L_j\) is the length of pipe \(j\).

The Darcy friction factor of a new pipe is provided for every option.
However, the rate at which the friction factor increases over time
(decay rate) is uncertain and is bounded between minimum and maximum
values.

Associated carbon emissions (in kg CO2-eq per meter of pipe installed)
are also provided. These emission factors may change over time due to
technological advancements.

Connections are either completely within a province (intra-province) or
shared between provinces (inter-province). The complete lists of
intra-province and inter-province connections are included in the data
files mapped in Appendix A. The capital cost for inter-provincial pipes
is shared equally between the water utilities operating in the connected
provinces (i.e., Equation~\ref{eq:pipes-capital-cost} is used but the
pipe's length is half).

\hypertarget{tbl:cn-properties}{}
\begin{longtable}[]{@{}llll@{}}
\caption{\label{tbl:cn-properties}Connections' properties
review.}\tabularnewline
\toprule\noalign{}
Property & Type & Scope & Unit \\
\midrule\noalign{}
\endfirsthead
\toprule\noalign{}
Property & Type & Scope & Unit \\
\midrule\noalign{}
\endhead
\bottomrule\noalign{}
\endlastfoot
Identifier & Static & Connection & \\
Node A & Static & Connection & \\
Node B & Static & Connection & \\
Type & Static & Connection & \\
Distance & Static & Connection & m \\
Pipe option & Decision & Connection & \\
Pipe installation date & Decision & Connection & \\
(Minor loss coefficient) & set to 0 & & \\
\end{longtable}

\hypertarget{tbl:pi-properties}{}
\begin{longtable}[]{@{}
  >{\raggedright\arraybackslash}p{(\columnwidth - 6\tabcolsep) * \real{0.2500}}
  >{\raggedright\arraybackslash}p{(\columnwidth - 6\tabcolsep) * \real{0.2500}}
  >{\raggedright\arraybackslash}p{(\columnwidth - 6\tabcolsep) * \real{0.2500}}
  >{\raggedright\arraybackslash}p{(\columnwidth - 6\tabcolsep) * \real{0.2500}}@{}}
\caption{\label{tbl:pi-properties}Pipes' properties
review.}\tabularnewline
\toprule\noalign{}
\begin{minipage}[b]{\linewidth}\raggedright
Property
\end{minipage} & \begin{minipage}[b]{\linewidth}\raggedright
Type
\end{minipage} & \begin{minipage}[b]{\linewidth}\raggedright
Scope
\end{minipage} & \begin{minipage}[b]{\linewidth}\raggedright
Unit
\end{minipage} \\
\midrule\noalign{}
\endfirsthead
\toprule\noalign{}
\begin{minipage}[b]{\linewidth}\raggedright
Property
\end{minipage} & \begin{minipage}[b]{\linewidth}\raggedright
Type
\end{minipage} & \begin{minipage}[b]{\linewidth}\raggedright
Scope
\end{minipage} & \begin{minipage}[b]{\linewidth}\raggedright
Unit
\end{minipage} \\
\midrule\noalign{}
\endhead
\bottomrule\noalign{}
\endlastfoot
Identifier & Static & Pipe option & \\
Diameter & Static & Pipe option & \\
Material & Static & Pipe option & \\
Darcy friction factor - new pipe & Static & Pipe option & \\
Darcy friction factor - decay rate & Static {[}Uncertain{]} & Pipe
option & \(years^-1\) \\
Darcy friction factor - existing pipe & Dynamic endogenous & Pipe option
& \\
Lifetime & Static {[}Uncertain{]} & Pipe option & \(years\) \\
Unit cost (new pipe) & Dynamic endogenous & Pipe option &
\(\text{€}/m\) \\
Equivalent emissions (new pipe) & Dynamic exogenous & Pipe option &
\(\text{tCO2eq}/m\) \\
\end{longtable}

\hypertarget{sec:ext-drivers}{%
\subsection{System External Drivers}\label{sec:ext-drivers}}

This chapter describes the external drivers that influence the national
WDS and how they have been modelled in the BWF competition. The key
external systems modelled are the climate, the energy system, and the
economic context.

These systems are beyond the direct control of the participants and
introduce the main exogenous factors shaping the environment in which
the transport network operates.

\hypertarget{climate-hydro-meteorological-forcing}{%
\subsubsection{Climate \& Hydro-meteorological
Forcing}\label{climate-hydro-meteorological-forcing}}

Climate is an exogenous factor beyond the control of the national water
sector. Four meteorological variables are relevant for the national grid
planning: temperature, precipitation, solar radiation and
evotranspiration.

The average maximum temperature affects water demand seasonality.

Temperature, precipitation, and evotranspiration influence river flows
(Meuse and Rhine). When these flows fall below mandated environmental
thresholds, surface water treatment plants must reduce operations or
close, temporarily limiting available supply.

All four variables affect groundwater availability and recharge rates,
potentially reducing the permitted extraction volumes of new sources in
the future. For a more detailed description of surface water and
groundwater source dynamics, see Section~\ref{sec:water-sources}.

Solar radiation influences the long-term electricity generation
potential of solar panels, as sustained higher irradiance levels lead to
greater cumulative energy yields over time.

The review of the climate variables can be seen in
Table~\ref{tbl:clim-properties}, while the actual values for these
variables can be inspected within the data files, which are mapped in
Appendix A. Note that instead of the evotranspiration variable, we
directly provide the Standardized Precipitation-Evapotranspiration Index
(SPEI) as it helps to quantify the severity, duration, and frequency of
droughts by indicating whether conditions are wetter or drier than
normal over a specific period.

\hypertarget{tbl:clim-properties}{}
\begin{longtable}[]{@{}llll@{}}
\caption{\label{tbl:clim-properties}Climate model's properties
review.}\tabularnewline
\toprule\noalign{}
Property & Type & Scope & Unit \\
\midrule\noalign{}
\endfirsthead
\toprule\noalign{}
Property & Type & Scope & Unit \\
\midrule\noalign{}
\endhead
\bottomrule\noalign{}
\endlastfoot
Temperature & Dynamic Exogenous & National & °C \\
Precipitation & Dynamic Exogenous & National & mm/day \\
Solar Radiation & Dynamic Exogenous & National & W/m² \\
Stand. Precip.-Evap. Ind. (SPEI) & Dynamic Exogenous & National & - \\
\end{longtable}

\hypertarget{sec:energy-model}{%
\subsubsection{Energy System}\label{sec:energy-model}}

The energy system in the BWF is modeled through three macro-drivers:
electricity prices, greenhouse gas emission factor, and solar panels
prices.

The electricity price is assumed to vary annually, influenced by
inflation but subject to high uncertainty. On a higher resolution,
prices also fluctuate hourly throughout the day. The electricity
purchased by the grid has an associated greenhouse gas emissions factor,
which can not be influcenced by the utilities but is adjusted yearly to
account for the evolution of the national electricity generation mix.

As an alternative to grid electricity, participants may install solar
photovoltaic panels at water sources or at pumping stations. The unit
cost of solar energy is modeled as a dynamic variable, changing annually
to reflect improvements in the technology.

The capital investment associated with the installation of solar panels
in water utility \(w\) at year \(y\) is:

\begin{equation}\protect\hypertarget{eq:solar-capital-investment}{}{
\text{CAPEX}^\text{solar}_w(y)
= \sum_{v \in \mathcal{V}_{w}} \mathbf{1}_{\{\tau_v = y\}}\cdot c^\text{solar}(y) \cdot P_v
}\label{eq:solar-capital-investment}\end{equation}

where \(v\) is the solar installation index within the set of
photovoltaic systems in the water utility (\(\mathcal{V}_{w}\)),
\(\tau_v\) is the installation time, \(\mathbf{1}_{\{\tau_v = y\}}\) is
an indicator function equal to 1 if the installation happened in year
\(y\) (0 otherwise), \(c^\text{solar}(y)\) is the unit cost per kilowatt
of capacity in year \(y\), and \(P_v\) is the nominal power capacity of
the solar system \(v\).

No cost is associated with the operation of the solar panels. These
components have a fixed lifespan of 25 years, after which competitors
must choose between investing in a replacement installation or reverting
to full dependence on the electrical grid.

\hypertarget{tbl:es-properties}{}
\begin{longtable}[]{@{}
  >{\raggedright\arraybackslash}p{(\columnwidth - 6\tabcolsep) * \real{0.2500}}
  >{\raggedright\arraybackslash}p{(\columnwidth - 6\tabcolsep) * \real{0.2500}}
  >{\raggedright\arraybackslash}p{(\columnwidth - 6\tabcolsep) * \real{0.2500}}
  >{\raggedright\arraybackslash}p{(\columnwidth - 6\tabcolsep) * \real{0.2500}}@{}}
\caption{\label{tbl:es-properties}Energy system model's properties
review.}\tabularnewline
\toprule\noalign{}
\begin{minipage}[b]{\linewidth}\raggedright
Property
\end{minipage} & \begin{minipage}[b]{\linewidth}\raggedright
Type
\end{minipage} & \begin{minipage}[b]{\linewidth}\raggedright
Scope
\end{minipage} & \begin{minipage}[b]{\linewidth}\raggedright
Unit
\end{minipage} \\
\midrule\noalign{}
\endfirsthead
\toprule\noalign{}
\begin{minipage}[b]{\linewidth}\raggedright
Property
\end{minipage} & \begin{minipage}[b]{\linewidth}\raggedright
Type
\end{minipage} & \begin{minipage}[b]{\linewidth}\raggedright
Scope
\end{minipage} & \begin{minipage}[b]{\linewidth}\raggedright
Unit
\end{minipage} \\
\midrule\noalign{}
\endhead
\bottomrule\noalign{}
\endlastfoot
Electricity price & Dynamic Exogenous & National & \(\text{€}/kWh\) \\
Emission factor & Dynamic Exogenous & National & \(tCO2eq/kWh\) \\
Unit cost - solar panel & Dynamic Endogenous & National &
\(\text{€}/kW\) \\
\end{longtable}

\hypertarget{economy}{%
\subsubsection{Economy}\label{economy}}

\hypertarget{inflation-dynamics}{%
\paragraph{Inflation Dynamics}\label{inflation-dynamics}}

Inflation captures the year-over-year change in the general price level.
In the BWF, it affects all costs equally except energy prices, which
follow different dynamics see \ref{sec:energy-model}.

As inflation compounds over time, it can substantially erode water
utilities' purchasing power and alter the financial viability of
capital-intensive interventions. While inflation is an exogenous
macroeconomic variable beyond the direct control of water utilities,
competitors can adopt strategic responses: front-loading major
investments to lock in current prices, or stress-testing their
masterplan against multiple inflation scenarios to ensure robustness.

In the absence of reliable long-term inflation forecasts, a reasonable
baseline assumption is the central bank's target rate (e.g., 2\% for the
European Central Bank).

\hypertarget{tbl:ei-properties}{}
\begin{longtable}[]{@{}llll@{}}
\caption{\label{tbl:ei-properties}Inflation's properties
review.}\tabularnewline
\toprule\noalign{}
Property & Type & Scope & Unit \\
\midrule\noalign{}
\endfirsthead
\toprule\noalign{}
Property & Type & Scope & Unit \\
\midrule\noalign{}
\endhead
\bottomrule\noalign{}
\endlastfoot
Inflation rate & Dynamic Exogenous & National & \\
\end{longtable}

Note that wherever unit prices are given only in base-year terms (e.g.,
only the value in year 2000 is reported in the input files), their value
in year \(y\) (\(C(y)\)) is obtained by multiplying by the cumulative
inflation index

\begin{equation}\protect\hypertarget{eq:interest-rate-prod}{}{
C(y) = C(0) \cdot \prod_{\tau = 1}^{y} \bigl(1 + \pi(\tau)\bigr)
}\label{eq:interest-rate-prod}\end{equation}

where \(C(0)\) is the base-year unit price and \(\pi(y)\) is the
inflation rate in year \(y\).

\hypertarget{budget-allocation}{%
\paragraph{Budget Allocation}\label{budget-allocation}}

A national budget (predefined over the planning horizon) is distributed
annually across utilities to fund infrastructure interventions and
operational expenses.

Competitors must decide how to distribute this budget among utilities in
every year. Budget distribution can follow several principled rules:
proportional allocation to population size; inverse-proportional schemes
favuoring smaller regions; income-based allocations reflecting
socioeconomic considerations; inverse-proportional schemes favuoring
less wealthy regions; or fully customised schemes.

\hypertarget{tbl:ea-properties}{}
\begin{longtable}[]{@{}llll@{}}
\caption{\label{tbl:ea-properties}Budget allocation model's properties
review.}\tabularnewline
\toprule\noalign{}
Property & Type & Scope & Unit \\
\midrule\noalign{}
\endfirsthead
\toprule\noalign{}
Property & Type & Scope & Unit \\
\midrule\noalign{}
\endhead
\bottomrule\noalign{}
\endlastfoot
National total budget & Static & National & €/year \\
Budget allocation policy & Option & Water utility & \\
\end{longtable}

\hypertarget{sec:water-pricing}{%
\paragraph{Water Pricing}\label{sec:water-pricing}}

Water utilities generate revenue selling water to three types of
customers: residential users, businesses, and, possibly, other water
utilities. Residential and commercial customers are billed using a
two-part pricing scheme: a fixed service charge (\(\text{€}/year\)) and
a volumetric charge based on consumption (\(\text{€}/m^3\)). For
simplicity, both retail customer types face identical rates within each
utility (with also no differentiation by income class or other
categories). Water transactions between utilities, however, use only a
volumetric charge based on the net exchange at the end of the year.

Thus, the total revenue for a water utility \(w\) in year \(y\) is:

\begin{equation}\protect\hypertarget{eq:revenue-water-utility}{}{
\text{REV}_w(y) = \sum_{m \in \mathcal{M}_w} P_w^\text{fixed}(y) + P_w^\text{variable}(y) \cdot Q^\text{BIL}_m(y) + \sum_{w' \in \mathcal{W}^-} P_w^\text{sell}(y) \cdot Q^{w'+}_w(y)
}\label{eq:revenue-water-utility}\end{equation}

where \(P_w^\text{fixed}(y)\) and \(P_w^\text{variable}(y)\) are the
fixed and volumetric components of the retail water price,
\(Q^\text{BIL}_m\) is the delivered billable demand in municipality
\(m\), \(\mathcal{M}_w\) is the set of municipalities served by water
utility \(w\), \(P_w^\text{sell}(y)\) is the volumetric charge that
utility \(w\) applies for inter-utility water sales, \(Q^{w'+}_w(y)\) is
the net positive volume of water sold by utility \(w\) to utility
\(w'\), and \(\mathcal{W}^-\) is the set of water utilities excluding
utility \(w\).

The net water exchange between utilities \(w\) and \(w'\) is defined as:

\begin{equation}\protect\hypertarget{eq:inter-utility-exchange}{}{
\begin{aligned}
\Delta Q_{w}^{w'}(y) &= \sum_{t \in \mathcal{Y}} \sum_{j \in \mathcal J _ {(w,w')}} Q_{j}(t) \\
Q_{w}^{w'+}(y) &= \max ( \Delta Q_{w}^{w'}(y), 0 ) \\
Q_{w}^{w'-}(y) &= - \min ( \Delta Q_{w}^{w'}(y), 0 )
\end{aligned}
}\label{eq:inter-utility-exchange}\end{equation}

where \(t\) is the simulation timestep, \(\mathcal{Y}\) is the set of
timesteps in year \(y\), \(Q_j(t)\) the flow over connection \(j\) from
the set of connections between the water utilities
\(\mathcal J_{(w,w')}\) (with flow direction positive from \(w\) to
\(w'\)).

Note that if a water utility has a negative net exchange with another
utility (i.e., \(\Delta Q_{w}^{w'}(y) <0\)), that will be regarded as a
water import cost:

\begin{equation}\protect\hypertarget{eq:water-purchase-utility}{}{
\text{WIC}_w(y) = \sum_{w' \in \mathcal{W}^-} P_{w'}^\text{sell}(y) \cdot Q^{w'-}_w(y)
}\label{eq:water-purchase-utility}\end{equation}

where \(P_{w'}^\text{sell}(y)\) is the volumetric charge applied by
utility \(w'\) for water sales.

Participants must decide the water pricing adjustment strategy. They can
either let all three quantities adjust according to inflation, or define
a custom policy by specifying the yearly percentage increase for each of
them independently.

\hypertarget{tbl:et-properties}{}
\begin{longtable}[]{@{}
  >{\raggedright\arraybackslash}p{(\columnwidth - 6\tabcolsep) * \real{0.2500}}
  >{\raggedright\arraybackslash}p{(\columnwidth - 6\tabcolsep) * \real{0.2500}}
  >{\raggedright\arraybackslash}p{(\columnwidth - 6\tabcolsep) * \real{0.2500}}
  >{\raggedright\arraybackslash}p{(\columnwidth - 6\tabcolsep) * \real{0.2500}}@{}}
\caption{\label{tbl:et-properties}Water price model's properties
review.}\tabularnewline
\toprule\noalign{}
\begin{minipage}[b]{\linewidth}\raggedright
Property
\end{minipage} & \begin{minipage}[b]{\linewidth}\raggedright
Type
\end{minipage} & \begin{minipage}[b]{\linewidth}\raggedright
Scope
\end{minipage} & \begin{minipage}[b]{\linewidth}\raggedright
Unit
\end{minipage} \\
\midrule\noalign{}
\endfirsthead
\toprule\noalign{}
\begin{minipage}[b]{\linewidth}\raggedright
Property
\end{minipage} & \begin{minipage}[b]{\linewidth}\raggedright
Type
\end{minipage} & \begin{minipage}[b]{\linewidth}\raggedright
Scope
\end{minipage} & \begin{minipage}[b]{\linewidth}\raggedright
Unit
\end{minipage} \\
\midrule\noalign{}
\endhead
\bottomrule\noalign{}
\endlastfoot
Fixed tariff component (service charge) & Dynamic Endogenous & Water
utility & €/year \\
Volumetric tariff component & Dynamic Endogenous & Water utility &
\(\text{€}/m^3\) \\
Water price for other utilities & Dynamic Endogenous & Water utility &
\(\text{€}/m^3\) \\
Water pricing adjustment policy & Option & Water utility & \\
\end{longtable}

\hypertarget{sec:bonds}{%
\paragraph{Bond Issuance}\label{sec:bonds}}

Whenever a water utility is unable to cover its expenditures in a given
year, it finances the resulting deficit by issuing nationally backed
bonds. Bonds are automatically generated to cover the utility debt in
that year.

Specifically, the bond amount is \(\kappa\) times the debt, i.e.,
\(\mathrm{amount}_i=\kappa \cdot \mathrm{debt}_w(y)\) where
\(\kappa \in [1,2.5]\). A value of \(\kappa\) equal to 1, implies that,
depending on investor demand and prevailing market conditions, the
proceeds from the bond issuance may barely cover the utility's financing
needs, leaving the balance for the following year around 0. In contrast,
higher values of \(\kappa\) (closer to 2.5) generate a cash surplus with
the downside of a larger principal obligation to be repaid at maturity.
Bonds are also characterised by a maturity of \(M\) years, determining
when the bond principal must be repaid, a coupon rate
(\(\mathrm{coupon}_i\)), which determines the interest payments due each
year, and a yield to maturity (\(\mathrm{yield}_i\)), which determines
the price of the bond (\(\mathrm{price}_i\)).

Each year, the utility must repay the sum of principal amounts of all
bonds reaching maturity plus the annual interest payments and and
receives proceeds based on the bond price in that year. Formally:

\begin{equation}\protect\hypertarget{eq:bonds-payment}{}{
\begin{aligned}
&\text{PRI}_w(y) = \sum_{i \in \mathcal {B}_w(y) : y=\tau_i+M} \mathrm{amount}_i \\
&\text{INT}_w(y) = \sum_{i \in \mathcal{B}_w(y)} \mathrm{amount}_i \cdot \mathrm{coupon}_i \\
&\mathrm{PRO}_w(y) = \mathrm{price}_i/100 \cdot \mathrm{amount}_i
\end{aligned}
}\label{eq:bonds-payment}\end{equation}

where \(i\) indicates the \emph{i}-th bond, \(\tau_i\) is the issuance
year for bond \(i\), \(\mathcal{B}_w(y)\) is the set of bonds active for
water utility \(w\) in year \(y\) and \(\mathcal {B}_w(y) : y=\tau_i+M\)
the subset of bonds reaching maturity \(M\).

The i-th bond's coupon, yield, and price are given by:

\begin{equation}\protect\hypertarget{eq:coupons-price}{}{
\begin{aligned}
&\mathrm{coupon}_i=r_f + \hat{\pi}(y=\tau_i) \\
&\mathrm{yield}_i=\mathrm{coupon}_i + a \cdot (1-d(y=\tau_i)) \\
&\mathrm{price}_i=\sum_{y=1}^M \frac{\mathrm{coupon}_i}{(1+\mathrm{yield}_i)^y} + \frac{100}{(1+\mathrm{yield}_i)^M}
\end{aligned}
}\label{eq:coupons-price}\end{equation}

where \(r_f\)\hspace{0pt} is the risk-free rate (long-term government
yield), \(\hat{\pi}(y=\tau_i)\) is the inflationary expectation at
issuance year, \(a\) is the sensitivity to investor demand, and
\(d(y=\tau_i)\)\hspace{0pt} is the uncertain demand factor for bond
\(i\) at issuance year.

Strong investor demand (\(d(y) > 1.0\)) increases \(\mathrm{PRO}_w(y)\)
(through a lower yield and therefore higher price at issuance), while
weak demand (\(d(y) < 1.0\)) decreases it. This simulates real-world
bond pricing where investor appetite introduces uncertainty to the
utilities budgetting.

While utilities cannot directly control bond yields, they can anticipate
debt accumulation through scenario analysis and adopt strategies that
maintain financial sustainability. For instance, utilities may design
pacing of interventions to minimize borrowing, or evaluate alternative
interventions that reduce the likelihood of large bond issuances during
periods with unfavorable yield conditions.

The complete list of the bond model properties can be seen in
Table~\ref{tbl:eb-properties}, while the actual values for these
variables can be inspected within the data files, which are mapped in
Appendix A.

\hypertarget{tbl:eb-properties}{}
\begin{longtable}[]{@{}llll@{}}
\caption{\label{tbl:eb-properties}Bonds model's properties
review.}\tabularnewline
\toprule\noalign{}
Property & Type & Scope & Unit \\
\midrule\noalign{}
\endfirsthead
\toprule\noalign{}
Property & Type & Scope & Unit \\
\midrule\noalign{}
\endhead
\bottomrule\noalign{}
\endlastfoot
Balance & Dynamic Endogenous & Water utility & € \\
Bond amount to debt ratio & Option & National & \\
Bond amount & Static & Bond & € \\
Bond issue date & Static & Bond & \\
Bond maturity & Static & National & years \\
Risk free rate & Static & National & \\
Inflationary expectations & Dynamic Exogenous & National & \\
Bond coupon & Dynamic Endogenous & Bond & \\
Sensitivity & Static & National & \\
Investors demand factor & Dynamic Exogenous & National & \\
Bond yield & Dynamic Endogenous & Bond & \\
Bond price & Dynamic Endogenous & Bond & \\
\end{longtable}

\hypertarget{sec:system-levers}{%
\subsection{System Levers}\label{sec:system-levers}}

Participants make two types of strategic decisions: policies and
interventions.

This chapter provides a summary of all the options available to
participants in the BWF competition (they have all been introduced in
previous sections already).

Here, we consolidate this information to give participants a complete
overview of the decision space they can navigate when developing their
masterplan.

\hypertarget{policies}{%
\subsubsection{Policies}\label{policies}}

Policies encompass regulatory and operational rules, such as pricing
structures, budget allocations, and maintenance protocols. Once set,
policies remain in effect until explicitly amended.

\hypertarget{sec:policy-budget-allocation}{%
\paragraph{National Budget Allocation
(National)}\label{sec:policy-budget-allocation}}

Participants must decide the strategy to allocate the national budget
across the water utilities. The policy can be a predefined one or follow
a custom allocation of the funds:

\begin{itemize}
\item
  \emph{by\_population}: Allocate the funds according to each province
  population.
\item
  \emph{by\_inverse\_population}: Allocate the funds to provinces with
  less population (less revenue).
\item
  \emph{by\_income}: Allocate the funds according to each province
  economic activity.
\item
  \emph{by\_inverse\_income}: Allocate the funds to provinces with less
  economic activity (less revenue).
\item
  \emph{custom}: Allocate the budget according to the specified share
  for each water utility (must sum to 1).
\end{itemize}

\begin{Shaded}
\begin{Highlighting}[]
\FunctionTok{year}\KeywordTok{:}\AttributeTok{ }\DecValTok{2025}
\AttributeTok{  }\FunctionTok{national\_policies}\KeywordTok{:}\AttributeTok{                               }
\AttributeTok{    }\FunctionTok{budget\_allocation}\KeywordTok{:}
\AttributeTok{      }\FunctionTok{policy}\KeywordTok{:}\AttributeTok{ by\_population}\CommentTok{ \# or "by\_income" or the corresponding inverse policies}
\end{Highlighting}
\end{Shaded}

\begin{Shaded}
\begin{Highlighting}[]
\FunctionTok{year}\KeywordTok{:}\AttributeTok{ }\DecValTok{2026}
\AttributeTok{  }\FunctionTok{national\_policies}\KeywordTok{:}\AttributeTok{                               }
\AttributeTok{    }\FunctionTok{budget\_allocation}\KeywordTok{:}
\AttributeTok{      }\FunctionTok{policy}\KeywordTok{:}\AttributeTok{ custom}
\AttributeTok{      }\FunctionTok{policy\_args}\KeywordTok{:}
\AttributeTok{        }\FunctionTok{WU01}\KeywordTok{:}\AttributeTok{ }\FloatTok{0.12}
\AttributeTok{        }\FunctionTok{WU02}\KeywordTok{:}\AttributeTok{ }\FloatTok{0.25}\CommentTok{ \# and so on... }
\end{Highlighting}
\end{Shaded}

\hypertarget{sec:policy-nrw-mitigation}{%
\paragraph{Non-Revenue Water Mitigation Budget
(Utility)}\label{sec:policy-nrw-mitigation}}

Participants must decide each water utility's yearly budget used to
mitigate non-revenue water (NRW). This budget is used to improve the
municipalities innner distribution network (IDN). More precisely, this
budget descreases the municipalities IDN's average age, which in turn
improves the NRW class of the municipality, leading to a reduction of
the NRW component. The policy can be a predefined one or follow a custom
allocation of the funds:

\begin{itemize}
\tightlist
\item
  \emph{by\_nrw\_class}: Allocate the budget to improve by one NRW-class
  each municipality in a greedy way (worst cases first) until no budget
  is left in that year.
\item
  \emph{by\_population}: Allocate the budget according to each
  municipality population.
\item
  \emph{custom}: Allocate the budget according to the specified share
  for each municipality (must sum to 1).
\end{itemize}

\begin{Shaded}
\begin{Highlighting}[]
\FunctionTok{year}\KeywordTok{:}\AttributeTok{ }\DecValTok{2025}
\AttributeTok{  }\FunctionTok{water\_utility}\KeywordTok{:}\AttributeTok{ WU01}
\AttributeTok{    }\FunctionTok{policies}\KeywordTok{:}
\AttributeTok{      }\FunctionTok{nrw\_mitigation}\KeywordTok{:}
\AttributeTok{        }\FunctionTok{budget}\KeywordTok{:}\AttributeTok{ }\DecValTok{30000}
\AttributeTok{        }\FunctionTok{policy}\KeywordTok{:}\AttributeTok{ by\_nrw\_class}\CommentTok{ \# or "by\_population"}
\end{Highlighting}
\end{Shaded}

\begin{Shaded}
\begin{Highlighting}[]
\FunctionTok{year}\KeywordTok{:}\AttributeTok{ }\DecValTok{2026}
\AttributeTok{  }\FunctionTok{water\_utility}\KeywordTok{:}\AttributeTok{ WU01}
\AttributeTok{    }\FunctionTok{policies}\KeywordTok{:}
\AttributeTok{      }\FunctionTok{nrw\_mitigation}\KeywordTok{:}
\AttributeTok{        }\FunctionTok{budget}\KeywordTok{:}\AttributeTok{ }\DecValTok{30000}
\AttributeTok{        }\FunctionTok{policy}\KeywordTok{:}\AttributeTok{ custom}
\AttributeTok{        }\FunctionTok{policy\_args}\KeywordTok{:}
\AttributeTok{          }\FunctionTok{GM0001}\KeywordTok{:}\AttributeTok{ }\FloatTok{0.02}
\AttributeTok{          }\FunctionTok{GM0002}\KeywordTok{:}\AttributeTok{ }\FloatTok{0.02}\CommentTok{ \# and so on...}
\end{Highlighting}
\end{Shaded}

\hypertarget{water-pricing-utility}{%
\paragraph{Water Pricing (Utility)}\label{water-pricing-utility}}

Participants must decide the water pricing adjustment strategy for each
year. They have two options: increase all water price components
according to inflation, or define a custom policy by specifying the
percentage increase for each component (e.g., 2\%).

\begin{Shaded}
\begin{Highlighting}[]
\FunctionTok{year}\KeywordTok{:}\AttributeTok{ }\DecValTok{2025}
\AttributeTok{  }\FunctionTok{water\_utility}\KeywordTok{:}\AttributeTok{ WU01}
\AttributeTok{    }\FunctionTok{policies}\KeywordTok{:}
\AttributeTok{      }\FunctionTok{pricing\_adjustment}\KeywordTok{:}
\AttributeTok{        }\FunctionTok{policy}\KeywordTok{:}\AttributeTok{ by\_inflation}
\end{Highlighting}
\end{Shaded}

\begin{Shaded}
\begin{Highlighting}[]
\FunctionTok{year}\KeywordTok{:}\AttributeTok{ }\DecValTok{2026}
\AttributeTok{  }\FunctionTok{water\_utility}\KeywordTok{:}\AttributeTok{ WU01}
\AttributeTok{    }\FunctionTok{policies}\KeywordTok{:}
\AttributeTok{      }\FunctionTok{pricing\_adjustment}\KeywordTok{:}
\AttributeTok{        }\FunctionTok{policy}\KeywordTok{:}\AttributeTok{ custom}
\CommentTok{        \# Specify the percentage increase for each pricing component}
\AttributeTok{        }\FunctionTok{policy\_args}\KeywordTok{:}
\AttributeTok{          }\FunctionTok{fixed\_component}\KeywordTok{:}\AttributeTok{ }\FloatTok{0.03}\CommentTok{      \# Annual increase for fixed costs (3\%)}
\AttributeTok{          }\FunctionTok{variable\_component}\KeywordTok{:}\AttributeTok{ }\FloatTok{0.02}\CommentTok{   \# Annual increase for variable costs (2\%)}
\AttributeTok{          }\FunctionTok{selling\_price}\KeywordTok{:}\AttributeTok{ }\FloatTok{0.05}\CommentTok{        \# Annual increase for water sales to other provinces (5\%) }
\end{Highlighting}
\end{Shaded}

\hypertarget{bond-issuance-utility}{%
\paragraph{Bond Issuance (Utility)}\label{bond-issuance-utility}}

Whenever the water utility is unable to cover its expenditures in a
specific year, it finances the resulting deficit by issuing nationally
backed bonds. Given that the raised amount is uncertain, participants
can cover this uncertainty increasing the bond amount ratio. This
adjustment is determined by the parameter \(\kappa\), which can be any
real number ranging from 1 to 2.5. More details are provided in
Section~\ref{sec:bonds}.

\begin{Shaded}
\begin{Highlighting}[]
\FunctionTok{year}\KeywordTok{:}\AttributeTok{ }\DecValTok{2025}
\AttributeTok{  }\FunctionTok{water\_utility}\KeywordTok{:}\AttributeTok{ WU01}
\AttributeTok{    }\FunctionTok{policies}\KeywordTok{:}
\AttributeTok{      }\FunctionTok{bond\_ratio}\KeywordTok{:}
\AttributeTok{        }\FunctionTok{value}\KeywordTok{:}\AttributeTok{ }\FloatTok{2.0}
\end{Highlighting}
\end{Shaded}

\hypertarget{interventions}{%
\subsubsection{Interventions}\label{interventions}}

Interventions are physical modifications to the system, such as
infrastructure upgrades or new installations. Interventions are
specified annually and are always implemented. Note that if a utility
contracts debt, a bond is issued automatically to cover the unbudgeted
expenses.

\hypertarget{opening-new-sources-utility}{%
\paragraph{Opening New Sources
(Utility)}\label{opening-new-sources-utility}}

Participants can open new water sources to meet potential increases in
demand. Available sources are predefined by location, and participants
must specify a capacity within the allowable bounds.

\begin{Shaded}
\begin{Highlighting}[]
\FunctionTok{year}\KeywordTok{:}\AttributeTok{ }\DecValTok{2025}
\AttributeTok{  }\FunctionTok{water\_utility}\KeywordTok{:}\AttributeTok{ WU01}
\AttributeTok{    }\FunctionTok{interventions}\KeywordTok{:}
\AttributeTok{      }\FunctionTok{open\_source}\KeywordTok{:}\CommentTok{ \# Provide source identifier, capacity, and info about the connection}
\AttributeTok{        }\KeywordTok{{-}}\AttributeTok{ }\FunctionTok{source\_id}\KeywordTok{:}\AttributeTok{ SG0158}
\AttributeTok{          }\FunctionTok{source\_capacity}\KeywordTok{:}\AttributeTok{ }\DecValTok{100}
\AttributeTok{          }\FunctionTok{pump\_option\_id}\KeywordTok{:}\AttributeTok{ PU003}
\AttributeTok{          }\FunctionTok{n\_pumps}\KeywordTok{:}\AttributeTok{ }\DecValTok{6}
\AttributeTok{          }\FunctionTok{pipe\_option\_id}\KeywordTok{:}\AttributeTok{ PI002}

\CommentTok{          \# Multiple sources can be added like this}
\AttributeTok{        }\KeywordTok{{-}}\AttributeTok{ }\FunctionTok{source\_id}\KeywordTok{:}\AttributeTok{ SG0159}
\AttributeTok{          }\FunctionTok{source\_capacity}\KeywordTok{:}\AttributeTok{ }\DecValTok{50}
\AttributeTok{          }\FunctionTok{pump\_option\_id}\KeywordTok{:}\AttributeTok{ PU001}
\AttributeTok{          }\FunctionTok{n\_pumps}\KeywordTok{:}\AttributeTok{ }\DecValTok{4}
\AttributeTok{          }\FunctionTok{pipe\_option\_id}\KeywordTok{:}\AttributeTok{ PI003}
\end{Highlighting}
\end{Shaded}

\hypertarget{closing-sources-utility}{%
\paragraph{Closing Sources (Utility)}\label{closing-sources-utility}}

Similarly, participants can close selected sources to improve the
overall system efficiency.

\begin{Shaded}
\begin{Highlighting}[]
\FunctionTok{year}\KeywordTok{:}\AttributeTok{ }\DecValTok{2025}
\AttributeTok{  }\FunctionTok{water\_utility}\KeywordTok{:}\AttributeTok{ WU01}
\AttributeTok{    }\FunctionTok{interventions}\KeywordTok{:}
\AttributeTok{      }\FunctionTok{close\_source}\KeywordTok{:}\CommentTok{ \# Provide only the source identifier}
\AttributeTok{        }\KeywordTok{{-}}\AttributeTok{ }\FunctionTok{source\_id}\KeywordTok{:}\AttributeTok{ SG0173    }

\CommentTok{          \# Multiple sources can be removed like this.}
\AttributeTok{        }\KeywordTok{{-}}\AttributeTok{ }\FunctionTok{source\_id}\KeywordTok{:}\AttributeTok{ SG0174}
\end{Highlighting}
\end{Shaded}

\hypertarget{installing-pipes-national-or-utility}{%
\paragraph{Installing Pipes (National or
Utility)}\label{installing-pipes-national-or-utility}}

Participants can decide to install new pipes or replace existing ones in
the system. Each installation requires specifying the connection
identifier and selecting a pipe option from a predefined list.

\begin{Shaded}
\begin{Highlighting}[]
\FunctionTok{year}\KeywordTok{:}\AttributeTok{ }\DecValTok{2025}
\AttributeTok{  }\FunctionTok{water\_utility}\KeywordTok{:}\AttributeTok{ WU01}
\AttributeTok{    }\FunctionTok{interventions}\KeywordTok{:}
\AttributeTok{      }\FunctionTok{install\_pipe}\KeywordTok{:}\CommentTok{ \# Provide connection identifier and pipe option}
\AttributeTok{        }\KeywordTok{{-}}\AttributeTok{ }\FunctionTok{connection\_id}\KeywordTok{:}\AttributeTok{ CG0112}
\AttributeTok{          }\FunctionTok{pipe\_option\_id}\KeywordTok{:}\AttributeTok{ PI001}

\CommentTok{          \# Multiple pipes can be added like this.             }
\AttributeTok{        }\KeywordTok{{-}}\AttributeTok{ }\FunctionTok{connection\_id}\KeywordTok{:}\AttributeTok{ CG0113}
\AttributeTok{          }\FunctionTok{pipe\_option\_id}\KeywordTok{:}\AttributeTok{ PI008}
\end{Highlighting}
\end{Shaded}

\hypertarget{installing-pumps-utility}{%
\paragraph{Installing Pumps (Utility)}\label{installing-pumps-utility}}

Participants can decide to install or replace pumps in pumping stations
to calibrate the peak outflow of water sources. Each source is
associated with one pumping station, which contains multiple identical
pumps operating in parallel. Pump options must be selected from a
predefined list. When installing pumps at an already-open pumping
station, participants must specify whether to replace or add to the
existing pumps. If the selected pump option differs from those already
installed, all existing pumps will be automatically replaced, as only
one pump type is allowed per station.

\begin{Shaded}
\begin{Highlighting}[]
\FunctionTok{year}\KeywordTok{:}\AttributeTok{ }\DecValTok{2025}
\AttributeTok{  }\FunctionTok{water\_utility}\KeywordTok{:}\AttributeTok{ WU01}
\AttributeTok{    }\FunctionTok{interventions}\KeywordTok{:}
\AttributeTok{      }\FunctionTok{install\_pumps}\KeywordTok{:}\CommentTok{ \# Provide source identifier, pump option and quantity for a new pumping station                                       }
\AttributeTok{        }\KeywordTok{{-}}\AttributeTok{ }\FunctionTok{source\_id}\KeywordTok{:}\AttributeTok{ SG0159}
\AttributeTok{          }\FunctionTok{pump\_option\_id}\KeywordTok{:}\AttributeTok{ PU003                                                             }
\AttributeTok{          }\FunctionTok{n\_pumps}\KeywordTok{:}\AttributeTok{ }\DecValTok{3}
\AttributeTok{          }\FunctionTok{behaviour}\KeywordTok{:}\AttributeTok{ replace}\CommentTok{ \# or "new" }
\end{Highlighting}
\end{Shaded}

\hypertarget{installing-solar-utility}{%
\paragraph{Installing Solar (Utility)}\label{installing-solar-utility}}

Participants can decide to install behind-the-meter solar panels at
water sources to reduce electricity costs and emissions for pumping
stations. The solar panels offset electricity consumption and emissions
but cannot be used as a profit-generating investment. Panels can be
installed multiple times at different points in time. \emph{Note: Solar
panels have a given lifespan (see Section~\ref{sec:energy-model});
participants must decide whether to replace them upon expiration.}

\begin{Shaded}
\begin{Highlighting}[]
\FunctionTok{year}\KeywordTok{:}\AttributeTok{ }\DecValTok{2025}
\AttributeTok{  }\FunctionTok{water\_utility}\KeywordTok{:}\AttributeTok{ WU01}
\AttributeTok{    }\FunctionTok{interventions}\KeywordTok{:}
\AttributeTok{      }\FunctionTok{install\_solar}\KeywordTok{:}\CommentTok{ \# Provide location and capacity}
\AttributeTok{        }\KeywordTok{{-}}\AttributeTok{ }\FunctionTok{source\_id}\KeywordTok{:}\AttributeTok{ SG0158}
\AttributeTok{          }\FunctionTok{capacity}\KeywordTok{:}\AttributeTok{ }\DecValTok{20}

\CommentTok{          \# Multiple sources can be added like this}
\AttributeTok{        }\KeywordTok{{-}}\AttributeTok{ }\FunctionTok{source\_id}\KeywordTok{:}\AttributeTok{ SG0159              }
\AttributeTok{          }\FunctionTok{capacity}\KeywordTok{:}\AttributeTok{ }\DecValTok{20}
\end{Highlighting}
\end{Shaded}

\hypertarget{system-requirements}{%
\subsection{System Requirements}\label{system-requirements}}

This chapter describes the metrics used to evaluate each solution
(masterplan) and the overall system performance in the BWF competition.
These metrics define the many-objective space that participants must
navigate, covering the main aspects of economic performance,
environmental impact, reliability, and fairness. Importantly, these are
evaluation metrics rather than explicit objectives: the competition
ranking formula will not be disclosed to participants. The ranking may
use weighted aggregation (with weights potentially changing over time to
reflect shifting social priorities) or custom multi-criteria evaluation
methods (e.g., selecting only solutions above a reliability threshold
before ranking by other criteria). However, participants may choose to
treat these metrics as objectives in their own optimization processes
when developing their masterplans.

\hypertarget{maintain-financial-viability}{%
\subsubsection{Maintain Financial
Viability}\label{maintain-financial-viability}}

Participants must seek to maintain all water utilities in a financially
viable state throughout the planning horizon. While utilities should not
pursue profit, they must avoid insolvency.

Each year (\(y\)), the financial balance \(F_w\) of every water utility
\(w\) is updated based on the following set of equations:

\begin{equation}\protect\hypertarget{eq:water-utility-finance}{}{
\begin{aligned}
F^*_{w}(y+1) &= F_{w}(y) + \text{NIB}(y) \cdot \alpha_w(y) + \text{REV}_{w}(y) \\
&\quad - \text{CAPEX}_w(y) - \text{OPEX}_w(y) \\
&\quad - \text{WLR}_w(y) -\text{WIC}_w(y) - \text{FIN}_w(y) \\
&\quad - \text{INT}_w(y) - \text{PRI}_w(y)\\
\text{debt}_w(y) &= \begin{cases}
-F_{w}^{*}(y+1) & \text{if } F_{w}^{*}(y+1) < 0 \\
0 & \text{otherwise}
\end{cases} \\
F_{w}(y+1) &= F_{w}^{*}(y+1) + \text{PRO}_w(y)
\end{aligned}
}\label{eq:water-utility-finance}\end{equation}

The provisional fund balance \(F_{w}^{*}(y+1)\) is calculated by
accounting for all inflows and outflows:

\begin{itemize}
\tightlist
\item
  \(\text{NIB}(y) \cdot \alpha_w(y)\) is the national investment budget
  allocated for the water utility \(w\) (see
  Section~\ref{sec:policy-budget-allocation}),
\item
  \(\text{REV}_w(y)\) is the water utility's revenue from the billable
  water demand (Equation~\ref{eq:revenue-water-utility}),
\item
  \(\text{CAPEX}_w(y)\) represents all the utility's interventions
  capital costs (i.e., the sum of
  Equations~\ref{eq:capex-sources}, \ref{eq:pumping-stations-capital-cost}, \ref{eq:pipes-capital-cost}, \ref{eq:solar-capital-investment}),
\item
  \(\text{OPEX}_w(y)\) accounts for all the utility's operational costs
  (i.e., the sum of
  Equations~\ref{eq:op-expends}, \ref{eq:pumping-stations-opex}),
\item
  \(\text{WLR}_w(y)\) is the budget for NRW mitigation (see
  Section~\ref{sec:policy-nrw-mitigation}),
\item
  \(\text{WIC}_w(y)\) is the cost for imported water from other water
  utilities (Equation~\ref{eq:water-purchase-utility}),
\item
  \(\text{FIN}_w(y)\) is the water displacement fine incurred due to
  groundwater overextraction
  (Table~\ref{tbl:gwsources-permit-violation}),
\item
  \(\text{INT}_w(y)\) the utility's interest payments due
  (Equation~\ref{eq:bonds-payment}), and
\item
  \(\text{PRI}_w(y)\) is the principal amount due
  (Equation~\ref{eq:bonds-payment}).
\end{itemize}

If the provisional balance is negative, the deficit defines the debt
(\(\text{debt}_w(y)\)), which triggers bond issuance. The bond proceeds
\(\text{PRO}_w(y)\) are determined according to Section~\ref{sec:bonds}
and Equations~\ref{eq:bonds-payment}, \ref{eq:coupons-price}.

The actual fund balance \(F_{w}(y+1)\) is obtained by adding the bond
proceeds (if any) to the provisional balance, ensuring the fund remains
solvent. Note that while surpluses are carried forward, deficits are
always financed through bond issuance automatically ensuring that the
fund balance is always positive.

Therefore, evaluation will focus only on the remaining debt at the end
of the planning period. Final debt measures insolvency risk, not
financial optimality. Competitors are free to pursue any financial
strategy they deem appropriate, such as minimising costs, adjusting
prices to enable expensive solutions, or adopting other innovative
approaches. The many-objective framework ensures balanced evaluation
across all dimensions. In other words, maintaining financial stability
is an achievable target; the real challenge is deciding what to
sacrifice.

\hypertarget{minimize-ghg-emissions}{%
\subsubsection{Minimize GHG Emissions}\label{minimize-ghg-emissions}}

Participants must keep the masterplan Greenhouse Gas (GHG) emissions to
the lowest feasible level.

The GHG emissions for water utility \(w\) in year \(y\) are:

\begin{equation}\protect\hypertarget{eq:ghg-emissions-calc}{}{
\mathrm{GHG}_w(y) = \mathrm{GHG}_w^{\text{emb}}(y)
+ \mathrm{GHG}_w^{\text{op}}(y)
}\label{eq:ghg-emissions-calc}\end{equation}

where \(\mathrm{GHG}_w^{\text{emb}}(y)\) the embedded (construction)
emissions, and \(\mathrm{GHG}_w^{\text{op}}(y)\) the operational
emissions from electricity use.

Only new pipes have embedded GHG emissions. The embedded emissions in
year \(y\) are:

\begin{equation}\protect\hypertarget{eq:ghg-emissions-pipes}{}{
\mathrm{GHG}_w^{\text{emb}}(y)
=  \sum_{c \in \mathcal{C}_w}
\mathbf{1}_{c\text{ activated in }y} \cdot
EF_{p_c}(y) \cdot L_c
}\label{eq:ghg-emissions-pipes}\end{equation}

where \(\mathcal{C}_w\) is the set of connections of water utility
\(w\)\footnote{as described in Section~\ref{sec:connections}, we
  distinguish between connections and pipes},
\(\mathbf{1}_{c\text{ activated in }y}\) is 1 if connection \(c\)
installs a new pipe in year \(y\), 0 otherwise, \(EF_{p_c}(y)\) is the
unit emission factor of the connection's selected pipe option \(p_c\),
and \(L_c\) is the connection length.

The unit emission factor \(EF_{p}\) for pipe option \(p\) depends on the
pipe option diameter \(Diam_p\) and material \(Mat_p\) and may change
over time because of technological advancements, i.e.,
\(EF_{p}(y) = EF(Diam_p, Mat_p, y)\).

The operational emissions are calculated based on the total electricity
purchased from the grid, covering both water treatment (sources) and
transport (pumping).

\begin{equation}\protect\hypertarget{eq:ghg-emissions-operational}{}{
\mathrm{GHG}_w^{\text{op}}(y) = \sum_{t \in \mathcal{Y}} \bigl[ \sum_{s \in \mathcal{S_w}} E_s(t) \cdot EF_s(t) + \sum_{p \in \mathcal{P}_w} E_p(t) \cdot EF_p(t) \bigr]
}\label{eq:ghg-emissions-operational}\end{equation}

where for each timestep \(t\) of a year \(y\)\footnote{\(y\) represent
  the year, while \(\mathcal{Y}\) is the collection of timesteps},
\(E_s(t)\) and \(EF_s(t)\) are the energy consumption and the emission
factor of source \(s\), while \(E_p(t)\) and \(EF_p(t)\) represent the
same quantities for pump \(p\).

Pumps energy consumption is retrieved via the EPANET simulations, while
the water sources energy consumption is calculated according to
Equation~\ref{eq:source-energy}\$.

The emission factors of both entities (pumping stations \(EF_p(t)\) and
sources \(EF_s(t)\)) are dynamic and depend on the size and time of
production of the behind-the-meter solar panels installation at that
location (if no solar is installed, this variable reduces to the
constant electricity grid emission factor in year \(y\), i.e.,
\(EF^{\text{el}}(y)\) ).

\hypertarget{maximize-service-reliability}{%
\subsubsection{Maximize Service
Reliability}\label{maximize-service-reliability}}

Participants must ensure high service reliability by minimizing unmet
water demand.

Service reliability for municipality \(m\) in year \(y\) is:

\begin{equation}\protect\hypertarget{eq:service-reliability}{}{
Rel_m(y) = 1 - \frac{U_m(y)}{D^{\text{BIL}}_m(y)}
}\label{eq:service-reliability}\end{equation}

where \(U_m(y)\) is the undelivered demand and \(D^{\text{BIL}}_m(y)\)
is the billable water demand.

Evaluation will focus on maintaining adequate service levels across all
municipalities of each water utility throughout the entire planning
horizon.

\hypertarget{promote-affordability-and-equity}{%
\subsubsection{Promote Affordability and
Equity}\label{promote-affordability-and-equity}}

Participants must ensure water remains affordable, particularly for
lower-income households, while maintaining equitable pricing across
municipalities.

The affordability fairness metric represents the fraction of income that
a household at the 20th percentile of the income distribution would
spend on essential water consumption. Affordability fairness (lower is
better) for water utility \(w\) in year \(y\) is:

\begin{equation}\protect\hypertarget{eq:affordability}{}{
AF_w(y) = \frac{P_w^{\text{fixed}}(y) + P_w^{\text{variable}}(y) \cdot D^{\text{life}}}{ADI_w^{p20}(y)}
}\label{eq:affordability}\end{equation}

where \(P_w^{\text{fixed}}(y)\) and \(P_w^{\text{variable}}(y)\) are the
fixed and variable components of water price, \(D^{\text{life}}\) is
lifeline volume (minimum water required per person), and
\(ADI_w^{p20}(y)\) is the 20th percentile of disposable income across
all households served by the water utility.

Evaluation will focus on minimizing affordability while maintaining
reasonable equity across utilities.

\hypertarget{acknowledgements}{%
\section*{Acknowledgements}\label{acknowledgements}}
\addcontentsline{toc}{section}{Acknowledgements}

We thank all KWR colleagues for their invaluable suggestions and support
throughout the project development. We would like to particularly thank
Mirjam Blokker, Mark Morley, Karel van Laarhoven, Djordje Mitrovic, Ina
Vertommen, Konstantinos Glynis, and Beatriz Gutierrez-Caloir.

Dennis, Lydia, Christos and Dragan would like to thank Yvonne
Hassink-Mulder at Vitens for her insights on real water utility
challenges; our collaboration profoundly shaped this battle from the
early stages.

We would like to thank Luke Butler, whose tool EPANET.js was
instrumental in developing this national network and enabled
visualization of the information with unprecedented ease.

Dennis, Lydia and Christos would like to thank their academic
supervisors Andrea Castelletti and Christos Makropoulos for their
support and guidance.

D. Zanutto, C. Michalopoulos, L. Tsiami, J. Brandt, D. Eliades, S.
Vrachimis, P. Samartzis, B. Hammer, P. Koundouri, M. Polycarpou, and D.
Savić have received funding from the European Research Council (ERC)
under the ERC Synergy Grant Water-Futures (Grant agreement No.~951424).

A. Artelt is supported by the Ministry of Culture and Science NRW
(Germany) as part of the Lamarr Fellow Network.

This publication reflects the views of the authors only.

\hypertarget{data-availability-statement}{%
\section*{Data Availability
Statement}\label{data-availability-statement}}
\addcontentsline{toc}{section}{Data Availability Statement}

All data versions are archived on Zenodo, with the latest version
serving as the official standard (Zanutto et al. 2025). The files are
also available in the
\href{https://github.com/WaterFutures/water-futures-battle/releases}{releases
section of the GitHub repository}.

The model uses data from trusted open sources under Creative Commons
licenses (e.g., CBS and KNMI) and draws on publicly available reports
(e.g., Vewin). Where not explicitly cited, values are based on the
authors' institutional knowledge or represent modeling assumptions.

A complete description of all data sources will be released upon
conclusion of the competition.

\hypertarget{notation}{%
\section*{Notation}\label{notation}}
\addcontentsline{toc}{section}{Notation}

\hypertarget{entity-identifiers}{%
\subsection*{Entity Identifiers}\label{entity-identifiers}}

Each entity is numbered sequentially starting from 1. Unique identifiers
consist of two capital letters indicating the object type, followed by a
number indicating the specific instance.

Jurisdictions:

\begin{itemize}
\tightlist
\item
  Nation: NL0000
\item
  Region (Landsdeel): LD0000
\item
  Province: PV0000
\item
  Municipality (Gemeente): GM0000
\end{itemize}

Water Sources:

\begin{itemize}
\tightlist
\item
  Groundwater: SG0000
\item
  Surface water: SS0000
\item
  Desalination: SD0000
\end{itemize}

Infrastructure:

\begin{itemize}
\tightlist
\item
  Pumping stations: PS0000 (uniquely determines the source)
\item
  Pump option: PU000
\end{itemize}

Connections:

\begin{itemize}
\tightlist
\item
  Between municipalities: CG0000
\item
  From source to municipality: CS0000
\item
  Between provinces: CP0000
\item
  Pipe option: PI000
\end{itemize}

\hypertarget{sets-and-indices}{%
\subsection*{Sets and Indices}\label{sets-and-indices}}

{[}Our indexes here{]}

\hypertarget{parameters}{%
\subsection*{Parameters}\label{parameters}}

{[}Our parameters here{]}

\hypertarget{decision-variables}{%
\subsection*{Decision Variables}\label{decision-variables}}

{[}Our decision variables here{]}

\hypertarget{references}{%
\section*{References}\label{references}}
\addcontentsline{toc}{section}{References}

\hypertarget{refs}{}
\begin{CSLReferences}{1}{0}
\leavevmode\vadjust pre{\hypertarget{ref-github:epanetplus}{}}%
Artelt, André. 2025. {``{EPANET-PLUS}.''} \emph{GitHub Repository}.
https://github.com/WaterFutures/EPANET-PLUS; GitHub.

\leavevmode\vadjust pre{\hypertarget{ref-wbhub}{}}%
Artelt, André, Katharina Giese, Stelios G. Vrachimis, Demetris G.
Eliades, Marios M. Polycarpou, and Barbara Hammer. 2025. {``{The
WaterBenchmarkHub: A Platform for Benchmarks in Water Distribution
Networks}.''} \emph{21st Computing and Control in the Water Industry
Conference (CCWI 2025) at the University of Sheffield (1st - 3rd
September 2025)}. \url{https://doi.org/10.15131/SHEF.DATA.29921051.V1}.

\leavevmode\vadjust pre{\hypertarget{ref-Basupi_Kapelan_2015}{}}%
Basupi, Innocent, and Zoran Kapelan. 2015. {``Flexible Water
Distribution System Design Under Future Demand Uncertainty.''}
\emph{Journal of Water Resources Planning and Management} 141 (4):
04014067. \url{https://doi.org/10.1061/(ASCE)WR.1943-5452.0000416}.

\leavevmode\vadjust pre{\hypertarget{ref-Beh_Maier_Dandy_2015}{}}%
Beh, Eva H. Y., Holger R. Maier, and Graeme C. Dandy. 2015. {``Adaptive,
Multiobjective Optimal Sequencing Approach for Urban Water Supply
Augmentation Under Deep Uncertainty.''} \emph{Water Resources Research}
51 (3): 1529--51. \url{https://doi.org/10.1002/2014WR016254}.

\leavevmode\vadjust pre{\hypertarget{ref-CBS_2025}{}}%
CBS. 2025. {``Homepage.''} \url{https://www.cbs.nl}.

\leavevmode\vadjust pre{\hypertarget{ref-Creaco_Franchini_Walski_2014}{}}%
Creaco, E., M. Franchini, and T. M. Walski. 2014. {``Accounting for
Phasing of Construction Within the Design of Water Distribution
Networks.''} \emph{Journal of Water Resources Planning and Management}
140 (5): 598--606.
\url{https://doi.org/10.1061/(ASCE)WR.1943-5452.0000358}.

\leavevmode\vadjust pre{\hypertarget{ref-Creaco_Franchini_Walski_2015}{}}%
---------. 2015. {``Taking Account of Uncertainty in Demand Growth When
Phasing the Construction of a Water Distribution Network.''}
\emph{Journal of Water Resources Planning and Management} 141 (2):
04014049. \url{https://doi.org/10.1061/(ASCE)WR.1943-5452.0000441}.

\leavevmode\vadjust pre{\hypertarget{ref-Haasnoot_Kwakkel_Walker_ter_Maat_2013}{}}%
Haasnoot, Marjolijn, Jan H. Kwakkel, Warren E. Walker, and Judith ter
Maat. 2013. {``Dynamic Adaptive Policy Pathways: A Method for Crafting
Robust Decisions for a Deeply Uncertain World.''} \emph{Global
Environmental Change} 23 (2): 485--98.
\url{https://doi.org/10.1016/j.gloenvcha.2012.12.006}.

\leavevmode\vadjust pre{\hypertarget{ref-Skerker_Zaniolo_Willebrand_Lickley_Fletcher_2023}{}}%
Skerker, J. B., M. Zaniolo, K. Willebrand, M. Lickley, and S. M.
Fletcher. 2023. {``Quantifying the Value of Learning for Flexible Water
Infrastructure Planning.''} \emph{Water Resources Research} 59 (6):
e2022WR034412. \url{https://doi.org/10.1029/2022WR034412}.

\leavevmode\vadjust pre{\hypertarget{ref-Vewin_2025}{}}%
Vewin. 2025. {``Homepage.''} \emph{Vewin}. \url{https://www.vewin.nl/}.

\leavevmode\vadjust pre{\hypertarget{ref-battle_zenodo}{}}%
Zanutto, Dennis, Christos Michalopoulos, Andrè Artelt, Stefano Alvisi,
Valentina Marsili, Filippo Mazzoni, and Panagiotis Samartzis. 2025.
{``The Battle of the Water Futures {[}Data and Software{]}.''} Zenodo.
\url{https://doi.org/10.5281/zenodo.17698299}.

\end{CSLReferences}

\newpage

\appendix

\hypertarget{appendix-a}{%
\section{Appendix A}\label{appendix-a}}

This appendix describes the data contained within the supplementary
material (zipped data folder). Four types of data are included in the
folder and they are described independently in the following sections.

\hypertarget{epanet-networks-files}{%
\subsection*{EPANET Networks Files}\label{epanet-networks-files}}

The provided network files were generated using EPANET 2.3.
Unfortunately, most common EPANET libraries and software (e.g., WNTR and
EPANET for Windows) are not keeping up with the versions and they may
generate warnings upon opening or fail to open them at all. When using a
library based on an outdated version of EPANET, the hydraulic results
will also differ, as EPANET 2.3 features numerous bug fixes with respect
to previous versions.

For these compatibility reason and to ensure consistency between
results, the solutions will be submitted through an ad-hoc form and inp
files will be rejected. The sole reference for correctness is the
evaluator's computations, which uses the latest EPANET 2.3 library.

At the time of writing, we suggest using
\href{https://epanetjs.com}{EPANET.js} to visualize the networks
overlaid on a map of the Netherlands, while for hydraulic computations,
we recommend using the EPANET-PLUS Python library, as it offers full
EPANET 2.3 support (Artelt 2025).

\hypertarget{configuration-file}{%
\subsection*{Configuration File}\label{configuration-file}}

The configuration file \texttt{configuration.yaml} tells the evaluator
how to build the system. Specifically, it indicates:

\begin{itemize}
\tightlist
\item
  The relative paths of the Excel input files describing the system
\item
  The parameters and settings that control the simulation
\item
  The interface to control which results to save
\end{itemize}

This makes the configuration file the central reference point that
connects all the input data sources with the simulation engine. By
modifying these parameters (e.g., switching input files), participants
can quickly simulate the system under different scenarios.

\hypertarget{settings}{%
\subsubsection*{Settings}\label{settings}}

\begin{itemize}
\tightlist
\item
  Start year (\texttt{start\_year}) {[} - {]}
\item
  End year (\texttt{end\_year}) {[} - {]}
\item
  National budget (\texttt{national\_budget}) {[}€{]}
\item
  Lifeline volume (\texttt{lifeline\_volume}) {[}L/day/person{]}
\end{itemize}

\hypertarget{evaluator-input-files}{%
\subsection*{Evaluator Input Files}\label{evaluator-input-files}}

The evaluator input files fully describe the system entities
(municipalities, sources, etc.) and modules (energy system, water demand
module, etc.). By modifying these files, participants can simulate
different scenarios and system configurations.

\hypertarget{static-properties-files}{%
\subsubsection*{Static Properties Files}\label{static-properties-files}}

Static Properties files define the characteristics of each entity, with
one entity per row and properties in columns. Multiple sheets can be
used to distinguish between subtypes of the same object (e.g., water
sources are divided into desalination, groundwater, and surface water).

\hypertarget{dynamic-properties-files}{%
\subsubsection*{Dynamic Properties
Files}\label{dynamic-properties-files}}

Dynamic Properties files contain time-varying data for both exogenous
inputs and initial values of endogenous variables. Each property has its
own sheet within the Excel file, organized as follows:

\begin{itemize}
\tightlist
\item
  Rows: Each row represents a snapshot in time, with the first column
  containing the timestamp
\item
  Columns: Each column represents a scope, i.e., which entity or group
  of entities the value applies to. For example, while `GMxxxx'
  indicates the value applies only to that specific municipality,
  `NL0000' indicates a national scope and therefore applies to all
  municipalities.
\end{itemize}

\textbf{When future values are reported} (e.g., timestamp after 2024 for
the first stage), \textbf{they can be regarded as expert-driven
scenarios.} Actual values will override these parameters in subsequent
stages of the competition.

\hypertarget{multi-parameter-static-and-dynamic-properties}{%
\subsubsection*{Multi-Parameter (Static and Dynamic)
Properties}\label{multi-parameter-static-and-dynamic-properties}}

When a property requires multiple parameters, column headers use a dash
separator (`-') to indicate:

\begin{itemize}
\tightlist
\item
  Uncertainty bounds:
  \texttt{{[}\textquotesingle{}NL0000-min\textquotesingle{},\ \textquotesingle{}NL0000-max\textquotesingle{}{]}}
\item
  Multi-dimensional scope:
  \texttt{{[}\textquotesingle{}A-SMALL\textquotesingle{},\ \textquotesingle{}A-MEDIUM\textquotesingle{},\ ...{]}}
  (e.g., variable varying by both class and size)
\item
  Time series patterns:
  \texttt{{[}\textquotesingle{}NL0000-1\textquotesingle{},\ \textquotesingle{}NL0000-2\textquotesingle{},\ ...{]}}
  (e.g., electricity price profiles at national level)
\end{itemize}

\hypertarget{modules}{%
\subsubsection*{Modules}\label{modules}}

\hypertarget{state}{%
\paragraph*{State}\label{state}}

\hypertarget{entities}{%
\subparagraph*{Entities}\label{entities}}

\begin{itemize}
\tightlist
\item
  Entity: \textbf{State}

  \begin{itemize}
  \tightlist
  \item
    File: \texttt{configuration.yaml}
  \item
    Properties:

    \begin{itemize}
    \tightlist
    \item
      State name (\texttt{name}) {[} - {]}
    \item
      State ID (\texttt{id}) {[} - {]}
    \end{itemize}
  \end{itemize}
\item
  Entity: \textbf{Region}

  \begin{itemize}
  \tightlist
  \item
    File: \texttt{jurisdictions/jurisdictions-static\_properties.xlsx}
  \item
    Sheet: \texttt{regions}
  \item
    Properties:

    \begin{itemize}
    \tightlist
    \item
      Region name (\texttt{name}) {[} - {]}
    \item
      Region ID (\texttt{cbs\_id}) {[} - {]}
    \item
      State ID (\texttt{state}) {[} - {]}
    \end{itemize}
  \end{itemize}
\item
  Entity: \textbf{Province}

  \begin{itemize}
  \tightlist
  \item
    File: \texttt{jurisdictions/jurisdictions-static\_properties.xlsx}
  \item
    Sheet: \texttt{provinces}
  \item
    Properties:

    \begin{itemize}
    \tightlist
    \item
      Province name (\texttt{name}) {[} - {]}
    \item
      Province ID (\texttt{cbs\_id}) {[} - {]}
    \item
      Region ID (\texttt{region}) {[} - {]}
    \end{itemize}
  \end{itemize}
\item
  Entity: \textbf{Municipality}

  \begin{itemize}
  \tightlist
  \item
    File: \texttt{jurisdictions/jurisdictions-static\_properties.xlsx}
  \item
    Sheet: \texttt{municipalities}
  \item
    Properties:

    \begin{itemize}
    \tightlist
    \item
      Municipality name (\texttt{name}) {[} - {]}
    \item
      Municipality ID (\texttt{cbs\_id}) {[} - {]}
    \item
      Province ID (\texttt{province}) {[} - {]}
    \item
      Opening date (\texttt{begin\_date}) {[} - {]}
    \item
      Closure date (\texttt{end\_date}) {[} - {]}
    \item
      Reason for the closure (\texttt{end\_reason}) {[} - {]}
    \item
      Municipality ID that inherits the hydraulic connections
      (\texttt{main\_destination\_municipality}) {[} - {]}
    \item
      Municipality IDs that acquired the municipality's assets
      (\texttt{destination\_cbs\_ids}) {[} - {]}
    \item
      Coordinates (\texttt{latitude}, \texttt{longitude},
      \texttt{elevation}) {[} degrees, degrees, m {]}
    \item
      Area characteristics (\texttt{touristic\_area\_cbs\_id},
      \texttt{touristic\_group\_cbs\_id}) {[} -, - {]}
    \end{itemize}
  \end{itemize}
\end{itemize}

\hypertarget{dynamic-properties}{%
\subparagraph*{Dynamic Properties}\label{dynamic-properties}}

\begin{itemize}
\tightlist
\item
  Property: \textbf{Population}

  \begin{itemize}
  \tightlist
  \item
    File: \texttt{jurisdictions/municipalities-dynamic\_properties.xlsx}
  \item
    Sheet: \texttt{population}
  \item
    Scope: Municipality ID, National
  \item
    Unit: {[} inhabitants {]}
  \item
    Notes: National values are for future projection
  \end{itemize}
\item
  Property: \textbf{Surface land}

  \begin{itemize}
  \tightlist
  \item
    File: \texttt{jurisdictions/municipalities-dynamic\_properties.xlsx}
  \item
    Sheet: \texttt{surface-land}
  \item
    Scope: Municipality ID
  \item
    Unit: {[} \(km^2\) {]}
  \end{itemize}
\item
  Property: \textbf{Surface water (inland)}

  \begin{itemize}
  \tightlist
  \item
    File: \texttt{jurisdictions/municipalities-dynamic\_properties.xlsx}
  \item
    Sheet: \texttt{surface-water-inland}
  \item
    Scope: Municipality ID
  \item
    Unit: {[} \(km^2\) {]}
  \end{itemize}
\item
  Property: \textbf{Surface water (open water)}

  \begin{itemize}
  \tightlist
  \item
    File: \texttt{jurisdictions/municipalities-dynamic\_properties.xlsx}
  \item
    Sheet: \texttt{surface-water-open}
  \item
    Scope: Municipality ID
  \item
    Unit: {[} \(km^2\) {]}
  \end{itemize}
\item
  Property: \textbf{Number of houses}

  \begin{itemize}
  \tightlist
  \item
    File: \texttt{jurisdictions/municipalities-dynamic\_properties.xlsx}
  \item
    Sheet: \texttt{n\_houses}
  \item
    Scope: Municipality ID
  \item
    Unit: {[} houses {]}
  \end{itemize}
\item
  Property: \textbf{Number of businesses}

  \begin{itemize}
  \tightlist
  \item
    File: \texttt{jurisdictions/municipalities-dynamic\_properties.xlsx}
  \item
    Sheet: \texttt{n\_businesses}
  \item
    Scope: Municipality ID
  \item
    Unit: {[} businesses {]}
  \end{itemize}
\item
  Property: \textbf{Associated water demand pattern - Residential}

  \begin{itemize}
  \tightlist
  \item
    File: \texttt{jurisdictions/municipalities-dynamic\_properties.xlsx}
  \item
    Sheet: \texttt{assoc\_dem\_pat-residential}
  \item
    Dimension: Pair
  \item
    Scope: Municipality ID
  \item
    Unit: {[} - {]}
  \end{itemize}
\item
  Property: \textbf{Associated water demand pattern - Business}

  \begin{itemize}
  \tightlist
  \item
    File: \texttt{jurisdictions/municipalities-dynamic\_properties.xlsx}
  \item
    Sheet: \texttt{assoc\_dem\_pat-business}
  \item
    Scope: Municipality ID
  \item
    Unit: {[} - {]}
  \end{itemize}
\item
  Property: \textbf{Average age of the Inner Distribution Network}

  \begin{itemize}
  \tightlist
  \item
    File: \texttt{jurisdictions/municipalities-dynamic\_properties.xlsx}
  \item
    Sheet: \texttt{dist\_network-age-avg}
  \item
    Scope: Municipality ID
  \item
    Unit: {[} years {]}
  \end{itemize}
\item
  Property: \textbf{Average disposable income}

  \begin{itemize}
  \tightlist
  \item
    File: \texttt{jurisdictions/municipalities-dynamic\_properties.xlsx}
  \item
    Sheet: \texttt{disposable\_income-avg}
  \item
    Scope: Municipality ID
  \item
    Unit: {[} k€ {]}
  \end{itemize}
\end{itemize}

\hypertarget{water-demand-model}{%
\paragraph*{Water Demand Model}\label{water-demand-model}}

\hypertarget{entities-1}{%
\subparagraph*{Entities}\label{entities-1}}

\begin{itemize}
\tightlist
\item
  Entity: \textbf{Water Demand Pattern}

  \begin{itemize}
  \tightlist
  \item
    File:
    \texttt{water\_demand\_model/water\_demand\_model-static\_properties.xlsx}
  \item
    Sheet: \texttt{residential}, \texttt{business}
  \item
    Properties:

    \begin{itemize}
    \tightlist
    \item
      Water Demand Pattern ID (\texttt{demand\_pattern\_id}) {[} - {]}
    \item
      Pattern values (\texttt{year\_hour}) {[} - {]} \emph{Dimension:
      hours of the year}
    \end{itemize}
  \end{itemize}
\end{itemize}

\hypertarget{dynamic-properties-1}{%
\subparagraph*{Dynamic Properties}\label{dynamic-properties-1}}

\begin{itemize}
\tightlist
\item
  Property: \textbf{Business Demand}

  \begin{itemize}
  \tightlist
  \item
    File:
    \texttt{water\_demand\_model/water\_demand\_model-dynamic\_properties.xlsx}
  \item
    Sheet: \texttt{per\_business\_demand}
  \item
    Scope: National
  \item
    Dimension: Uniform Uncertain
  \item
    Unit: {[} \(m^3/business/hour\) {]}
  \end{itemize}
\item
  Property: \textbf{House Demand}

  \begin{itemize}
  \tightlist
  \item
    File:
    \texttt{water\_demand\_model/water\_demand\_model-dynamic\_properties.xlsx}
  \item
    Sheet: \texttt{per\_house\_demand}
  \item
    Scope: National
  \item
    Dimension: Uniform Uncertain
  \item
    Unit: {[} \(m^3/house/hour\) {]}
  \end{itemize}
\end{itemize}

\hypertarget{non-revenue-water-model}{%
\paragraph*{Non-Revenue Water Model}\label{non-revenue-water-model}}

\hypertarget{static-properties}{%
\subparagraph*{Static Properties}\label{static-properties}}

\begin{itemize}
\tightlist
\item
  Property: \textbf{NRW Intervention success probability}

  \begin{itemize}
  \tightlist
  \item
    File: \texttt{configuration.yaml}
  \item
    Label: \texttt{nrw\_model-intervention\_success\_prob}
  \item
    Scope: National
  \item
    Dimension: Uniform Uncertain
  \item
    Unit: {[} - {]}
  \end{itemize}
\end{itemize}

\hypertarget{dynamic-properties-2}{%
\subparagraph*{Dynamic Properties}\label{dynamic-properties-2}}

\begin{itemize}
\tightlist
\item
  Property: \textbf{NRW Intervention Unit cost}

  \begin{itemize}
  \tightlist
  \item
    File: \texttt{jurisdictions/nrw\_model-dynamic\_properties.xlsx}
  \item
    Sheet: \texttt{nrw\_intervention-unit\_cost}
  \item
    Scope: National
  \item
    Dimension: NRW class \(\times\) Municipality size class
  \item
    Unit: {[} \(\text{€}/year/km\) {]}
  \end{itemize}
\end{itemize}

\hypertarget{sources}{%
\paragraph*{Sources}\label{sources}}

\hypertarget{entities-2}{%
\subparagraph*{Entities}\label{entities-2}}

\begin{itemize}
\tightlist
\item
  Entity: \textbf{Water Source} (Groundwater, Surface water,
  Desalination)

  \begin{itemize}
  \tightlist
  \item
    File: \texttt{sources/sources-static\_properties.xlsx}
  \item
    Sheet: \texttt{groundwater}, \texttt{surface\_water},
    \texttt{desalination}
  \item
    Properties:

    \begin{itemize}
    \tightlist
    \item
      Source Name (\texttt{name}) {[} - {]}
    \item
      Source ID (\texttt{source\_id}) {[} - {]}
    \item
      Coordinates (\texttt{latitude}, \texttt{longitude},
      \texttt{elevation}) {[} degrees, degrees, m {]}
    \item
      Province (\texttt{province}) {[} - {]}
    \item
      Closest municipality ID (\texttt{closest\_municipality})
    \item
      Activation date (\texttt{activation\_date}) {[} - {]}
    \item
      Closure date (\texttt{closure\_date}) {[} - {]}
    \item
      Nominal capacity (\texttt{capacity-nominal}) {[} \(m^3/day\) {]}
    \item
      Specific energy (\texttt{opex-volum-energy\_factor}) {[}
      \(kWh/m^3\) {]}
    \item
      Source permit (\texttt{permit}) {[} m\(m^3/year\) {]} \emph{Only
      groundwater}
    \item
      Basin (\texttt{basin}) {[} - {]} \emph{Only surface water}
    \end{itemize}
  \end{itemize}
\end{itemize}

\hypertarget{static-properties-1}{%
\subparagraph*{Static Properties}\label{static-properties-1}}

\begin{itemize}
\tightlist
\item
  Property: \textbf{Capacity Target Factor}

  \begin{itemize}
  \tightlist
  \item
    File: \texttt{sources/sources-static\_properties.xlsx}
  \item
    Sheet: \texttt{global}
  \item
    Label: \texttt{capacity-target\_factor}
  \item
    Scope: Source Type
  \item
    Unit {[} - {]}
  \end{itemize}
\item
  Property: \textbf{Operational costs - volumetric for non-energy -
  multiplier}

  \begin{itemize}
  \tightlist
  \item
    File: \texttt{sources/sources-static\_properties.xlsx}
  \item
    Sheet: \texttt{global}
  \item
    Label \texttt{opex-volum-other-multiplier}
  \item
    Scope: Source Type
  \item
    Unit {[} - {]}
  \end{itemize}
\item
  Property: \textbf{Construction time}

  \begin{itemize}
  \tightlist
  \item
    File: \texttt{sources/sources-static\_properties.xlsx}
  \item
    Sheet: \texttt{global}
  \item
    Scope: Source Type
  \item
    Dimension: Uniform Uncertain
  \item
    Unit {[} - {]}
  \end{itemize}
\item
  Property: \textbf{Operational costs - volumetric for energy (specific
  energy)}

  \begin{itemize}
  \tightlist
  \item
    File: \texttt{sources/sources-static\_properties.xlsx}
  \item
    Sheet: \texttt{global}
  \item
    Scope: Source Type
  \item
    Dimension: Uniform Uncertain
  \item
    Unit {[} - {]}
  \item
    Notes: sampled for new sources
  \end{itemize}
\end{itemize}

\hypertarget{dynamic-properties-3}{%
\subparagraph*{Dynamic Properties}\label{dynamic-properties-3}}

\begin{itemize}
\tightlist
\item
  Property: \textbf{Unit cost of construction}

  \begin{itemize}
  \tightlist
  \item
    File: \texttt{sources/\{source\_type\}-dynamic\_properties.xlsx}
  \item
    Sheet: \texttt{new\_source-unit\_cost}
  \item
    Scope: National
  \item
    Dimension: Source size
  \item
    Unit: {[} \(\text{€}/(m^3/day)\) {]}
  \end{itemize}
\item
  Property: \textbf{Operational costs - fixed (unit cost)}

  \begin{itemize}
  \tightlist
  \item
    File: \texttt{sources/\{source\_type\}-dynamic\_properties.xlsx}
  \item
    Sheet: \texttt{opex-fixed}
  \item
    Scope: National
  \item
    Dimension: Source size \(\times\) Uniform Uncertain
  \item
    Unit: {[} \(\text{€}/(m^3)\) {]}
  \item
    Notes: gets multiplied by annual nominal capacity
  \end{itemize}
\item
  Property: \textbf{Operational costs - volumetric for non-energy}

  \begin{itemize}
  \tightlist
  \item
    File: \texttt{sources/\{source\_type\}-dynamic\_properties.xlsx}
  \item
    Sheet: \texttt{opex-volum-other}
  \item
    Scope: National
  \item
    Dimension: Source size \(\times\) Uniform Uncertain
  \item
    Unit: {[} \(\text{€}/(m^3)\) {]}
  \end{itemize}
\item
  Property: \textbf{Availability factor}

  \begin{itemize}
  \tightlist
  \item
    File: \texttt{sources/\{source\_type\}-dynamic\_properties.xlsx}
  \item
    Sheet: \texttt{availability\_factor}
  \item
    Scope: National, Basin
  \item
    Unit: {[} - {]}
  \end{itemize}
\item
  Property: \textbf{Water displacement fine amount}

  \begin{itemize}
  \tightlist
  \item
    File: \texttt{sources/groundwater-dynamic\_properties.xlsx}
  \item
    Sheet: \texttt{water\_displacement-fine\_amount}
  \item
    Scope: National
  \item
    Dimension: Displacement severity
  \item
    Unit: {[} € {]}
  \end{itemize}
\end{itemize}

\hypertarget{pumping-infrastructure}{%
\paragraph*{Pumping Infrastructure}\label{pumping-infrastructure}}

\hypertarget{entities-3}{%
\subparagraph*{Entities}\label{entities-3}}

\begin{itemize}
\tightlist
\item
  Entity: \textbf{Pumping Station}

  \begin{itemize}
  \tightlist
  \item
    File:
    \texttt{pumping\_stations/pumping\_stations-static\_properties.xlsx}
  \item
    Sheet: \texttt{entities}
  \item
    Properties:

    \begin{itemize}
    \tightlist
    \item
      Pumping Station ID (\texttt{pumping\_station\_id}) {[} - {]}
    \item
      Assigned source (\texttt{assigned\_source}) {[} - {]}
    \item
      Installed pumps - options IDs (\texttt{pumps-option\_ids}) {[} -
      {]}
    \item
      Installed pumps - installation date
      (\texttt{pumps-installation\_dates}) {[} - {]}
    \item
      Installed pumps - decomission date (\texttt{pumps-end\_dates}) {[}
      - {]}
    \end{itemize}
  \end{itemize}
\item
  Entity: \textbf{Pump Options}

  \begin{itemize}
  \tightlist
  \item
    File: \texttt{pumps/pump\_options-static\_properties.xlsx}
  \item
    Sheet: \texttt{options}
  \item
    Properties:

    \begin{itemize}
    \tightlist
    \item
      Pump Option ID (\texttt{option\_id}) {[} - {]}
    \item
      Name (\texttt{name}) {[} - {]}
    \item
      Nominal Flow rate (\texttt{flow\_rate-nominal}) {[} \(m^3/hour\)
      {]}
    \item
      Lifetime (\texttt{lifetime}) {[} years {]} Dimension: Uniform
      Uncertain
    \end{itemize}
  \end{itemize}
\item
  Entity: \textbf{Pump Curve}

  \begin{itemize}
  \tightlist
  \item
    File: \texttt{pumps/pump\_options-static\_properties.xlsx}
  \item
    Sheet: \texttt{\{option\_id\}}
  \item
    Properties:

    \begin{itemize}
    \tightlist
    \item
      Flowrate (\texttt{flowrate}) {[} \(m^3/hour\) {]}
    \item
      Head (\texttt{head}) {[} m {]}
    \item
      Efficiency (\texttt{efficiency}) {[} - {]}
    \end{itemize}
  \end{itemize}
\end{itemize}

\hypertarget{dynamic-properties-4}{%
\subparagraph*{Dynamic Properties}\label{dynamic-properties-4}}

\begin{itemize}
\tightlist
\item
  Property: \textbf{Unit cost for a new pump}

  \begin{itemize}
  \tightlist
  \item
    File: \texttt{pumps/pump\_options-dynamic\_properties.xlsx}
  \item
    Sheet: \texttt{new\_pump-cost}
  \item
    Scope: Pump Option
  \item
    Unit: {[} € {]}
  \end{itemize}
\end{itemize}

\hypertarget{piping-infrastructure}{%
\paragraph*{Piping Infrastructure}\label{piping-infrastructure}}

\hypertarget{entities-4}{%
\subparagraph*{Entities}\label{entities-4}}

\begin{itemize}
\tightlist
\item
  Entity: \textbf{Connection}

  \begin{itemize}
  \tightlist
  \item
    File: \texttt{connections/connections-static\_properties.xlsx}
  \item
    Sheet: \texttt{provincial}, \texttt{sources},
    \texttt{cross-provincial}
  \item
    Properties:

    \begin{itemize}
    \tightlist
    \item
      Connection ID (\texttt{connection\_id}) {[} - {]}
    \item
      From node (\texttt{from\_node}) {[} - {]}
    \item
      To node (\texttt{to\_node}) {[} - {]}
    \item
      Distance (\texttt{distance}) {[} m {]}
    \item
      Minor loss coefficient (\texttt{minor\_loss\_coeff}) {[} - {]}
    \item
      Installed pipes - options IDs (\texttt{pipes-option\_ids}) {[} -
      {]}
    \item
      Installed pipes - installation date
      (\texttt{pipes-installation\_dates}) {[} - {]}
    \item
      Installed pipes - decommission date
      (\texttt{pipes-decommission\_dates}) {[} - {]}
    \item
      Replaced by connection ID (\texttt{replaced\_by}) {[} - {]}
    \item
      Replaced connection IDs (\texttt{replaces}) {[} - {]}
    \end{itemize}
  \end{itemize}
\item
  Entity: \textbf{Pipe Options}

  \begin{itemize}
  \tightlist
  \item
    File: \texttt{pipes/pipe\_options-static\_properties.xlsx}
  \item
    Sheet: \texttt{options}
  \item
    Properties:

    \begin{itemize}
    \tightlist
    \item
      Pipe Option ID (\texttt{option\_id}) {[} - {]}
    \item
      Diameter (\texttt{diameter}) {[} mm {]}
    \item
      Material (\texttt{material}) {[} - {]}
    \item
      Darcy friction factor - New pipe
      (\texttt{darcy\_friction\_factor-new\_pipe}) {[} - {]}
    \item
      Darcy friction factor - Decay rate
      (\texttt{darcy\_friction\_factor-decay\_rate}) {[} - {]}
      Dimension: Uniform Uncertain
    \item
      Lifetime (\texttt{lifetime}) {[} years {]} Dimension: Uniform
      Uncertain
    \end{itemize}
  \end{itemize}
\end{itemize}

\hypertarget{dynamic-properties-5}{%
\subparagraph*{Dynamic Properties}\label{dynamic-properties-5}}

\begin{itemize}
\tightlist
\item
  Property: \textbf{New Pipe Unit Cost}

  \begin{itemize}
  \tightlist
  \item
    File: \texttt{connections/pipe\_options-dynamic\_properties.xlsx}
  \item
    Sheet: \texttt{new\_pipe-unit\_cost}
  \item
    Unit: {[} \(\text{€}/m\) {]}
  \end{itemize}
\item
  Property: \textbf{New Pipe Emission Factor}

  \begin{itemize}
  \tightlist
  \item
    File: \texttt{connections/pipe\_options-dynamic\_properties.xlsx}
  \item
    Sheet: \texttt{new\_pipe-emissions\_factor}
  \item
    Unit: {[} \(\text{tCO2eq}/m\) {]}
  \end{itemize}
\end{itemize}

\hypertarget{climate}{%
\paragraph*{Climate}\label{climate}}

\hypertarget{dynamic-properties-6}{%
\subparagraph*{Dynamic Properties}\label{dynamic-properties-6}}

\begin{itemize}
\tightlist
\item
  Property: \textbf{Temperature} (Average, Average of minimum/maximum
  daily temperatures, Maximum/Minimum temperature recorded)

  \begin{itemize}
  \tightlist
  \item
    File: \texttt{climate/climate-dynamic\_properties.xlsx}
  \item
    Sheet: \texttt{temperature-avg},
    \texttt{temperature-min-avg},\texttt{temperature-max-avg},
    \texttt{temperature-warmest\_day}, \texttt{temperature-coldest\_day}
  \item
    Scope: National
  \item
    Unit: {[} \(\text{°C}\) {]}
  \end{itemize}
\item
  Property: \textbf{Precipitation}

  \begin{itemize}
  \tightlist
  \item
    File: \texttt{climate/climate-dynamic\_properties.xlsx}
  \item
    Sheet: \texttt{precipitation}
  \item
    Scope: National
  \item
    Unit: {[} \(\text{mm}/day\) {]}
  \end{itemize}
\item
  Property: \textbf{Solar Radiation}

  \begin{itemize}
  \tightlist
  \item
    File: \texttt{climate/climate-dynamic\_properties.xlsx}
  \item
    Sheet: \texttt{solar\_radiation-avg}
  \item
    Scope: National
  \item
    Unit: {[} \(\text{W}/m^2\) {]}
  \end{itemize}
\item
  Property: \textbf{Standardized Precipitation-Evotranspiration Index}

  \begin{itemize}
  \tightlist
  \item
    File: \texttt{climate/climate-dynamic\_properties.xlsx}
  \item
    Sheet: \texttt{SPEI}
  \item
    Scope: National
  \item
    Unit: {[} - {]}
  \end{itemize}
\end{itemize}

\hypertarget{economy-1}{%
\paragraph*{Economy}\label{economy-1}}

\hypertarget{entities-5}{%
\subparagraph*{Entities}\label{entities-5}}

\begin{itemize}
\tightlist
\item
  Entity: \textbf{Bonds}

  \begin{itemize}
  \tightlist
  \item
    File: \texttt{economy/bonds-static\_properties.xlsx}
  \item
    Sheet: \texttt{entities}
  \item
    Properties:

    \begin{itemize}
    \tightlist
    \item
      Bond issuance ID (\texttt{bond\_issuance\_id}) {[} - {]}
    \item
      Number of bonds (\texttt{n\_bonds}) {[} - {]}
    \item
      Issue date (\texttt{issue\_date}) {[} - {]}
    \item
      Maturity date (\texttt{maturity\_date}) {[} - {]}
    \item
      Coupon rate (\texttt{coupon\_rate}) {[} - {]}
    \item
      Water Utility ID (\texttt{water\_utility\_id}) {[} - {]}
    \end{itemize}
  \end{itemize}
\end{itemize}

\hypertarget{dynamic-properties-7}{%
\subparagraph*{Dynamic Properties}\label{dynamic-properties-7}}

\begin{itemize}
\tightlist
\item
  Property: \textbf{Inflation}

  \begin{itemize}
  \tightlist
  \item
    File: \texttt{economy/economy-dynamic\_properties.xlsx}
  \item
    Sheet: \texttt{inflation}
  \item
    Scope: National
  \item
    Unit: {[} \% {]}
  \end{itemize}
\item
  Property: \textbf{Inflation Expectation}

  \begin{itemize}
  \tightlist
  \item
    File: \texttt{economy/economy-dynamic\_properties.xlsx}
  \item
    Sheet: \texttt{inflation-expected}
  \item
    Scope: National
  \item
    Unit: {[} \% {]}
  \end{itemize}
\item
  Property: \textbf{Investor Demand}

  \begin{itemize}
  \tightlist
  \item
    File: \texttt{economy/economy-dynamic\_properties.xlsx}
  \item
    Sheet: \texttt{investor\_demand}
  \item
    Scope: National
  \item
    Unit: {[} - {]}
  \end{itemize}
\end{itemize}

\hypertarget{energy-system}{%
\paragraph*{Energy System}\label{energy-system}}

\hypertarget{entities-6}{%
\subparagraph*{Entities}\label{entities-6}}

\begin{itemize}
\tightlist
\item
  Entity: \textbf{Solar Farm}

  \begin{itemize}
  \tightlist
  \item
    File: \texttt{energy/solar\_farms-static\_properties.xlsx}
  \item
    Sheet: \texttt{entities}
  \item
    Properties:

    \begin{itemize}
    \tightlist
    \item
      Solar farm ID (\texttt{solar\_farm\_id}) {[} - {]}
    \item
      Capacity (\texttt{capacity}) {[} kW {]}
    \item
      Installation date (\texttt{installation\_date}) {[} - {]}
    \item
      Decommission date (\texttt{decommission\_date}) {[} - {]}
    \item
      Connected entity ID source/pumping station
      (\texttt{connected\_entity\_id}) {[} - {]}
    \end{itemize}
  \end{itemize}
\end{itemize}

\hypertarget{dynamic-properties-8}{%
\subparagraph*{Dynamic Properties}\label{dynamic-properties-8}}

\begin{itemize}
\tightlist
\item
  Property: \textbf{Electricity Price}

  \begin{itemize}
  \tightlist
  \item
    File: \texttt{energy/energy\_system-dynamic\_properties.xlsx}
  \item
    Sheet: \texttt{electricity\_price-unit\_cost}
  \item
    Scope: National
  \item
    Unit: {[} \(\text{€}/kWh\) {]}
  \end{itemize}
\item
  Property: \textbf{Electricity Price Pattern}

  \begin{itemize}
  \tightlist
  \item
    File: \texttt{energy/energy\_system-dynamic\_properties.xlsx}
  \item
    Sheet: \texttt{electricity\_price-pattern}
  \item
    Scope: National
  \item
    Dimension: Hour of the week
  \item
    Unit: {[} - {]}
  \end{itemize}
\item
  Property: \textbf{Grid Emission factor}

  \begin{itemize}
  \tightlist
  \item
    File: \texttt{energy/energy\_system-dynamic\_properties.xlsx}
  \item
    Sheet: \texttt{grid\_emission\_factor}
  \item
    Scope: National
  \item
    Unit: {[} \(kgCO_2eq/kWh\) {]}
  \end{itemize}
\item
  Property: \textbf{Solar Panel Unit Cost}

  \begin{itemize}
  \tightlist
  \item
    File: \texttt{energy/energy\_system-dynamic\_properties.xlsx}
  \item
    Sheet: \texttt{solar\_panel-unit\_cost}
  \item
    Scope: National
  \item
    Unit: {[} \(\text{€}/kW\) {]}
  \end{itemize}
\end{itemize}

\hypertarget{water-utilities-1}{%
\paragraph*{Water Utilities}\label{water-utilities-1}}

\hypertarget{entities-7}{%
\subparagraph*{Entities}\label{entities-7}}

\begin{itemize}
\tightlist
\item
  Entity: \textbf{Water Utility}

  \begin{itemize}
  \tightlist
  \item
    File:
    \texttt{water\_utilities/water\_utilities-static\_properties.xlsx}
  \item
    Sheet: \texttt{entities}
  \item
    Properties:

    \begin{itemize}
    \tightlist
    \item
      Identifier (\texttt{water\_utility\_id}) {[} - {]}
    \item
      Assigned provinces (\texttt{assigned\_provinces}) {[} - {]}
    \end{itemize}
  \end{itemize}
\end{itemize}

\hypertarget{dynamic-properties-9}{%
\subparagraph*{Dynamic Properties}\label{dynamic-properties-9}}

\begin{itemize}
\tightlist
\item
  Property: \textbf{Funds Balance}

  \begin{itemize}
  \tightlist
  \item
    File:
    \texttt{water\_utilities/water\_utilities-dynamic\_properties.xlsx}
  \item
    Sheet: \texttt{balance}
  \item
    Scope: Water Utility
  \item
    Unit: {[} € {]}
  \end{itemize}
\item
  Property: \textbf{Water Price Fixed Component}

  \begin{itemize}
  \tightlist
  \item
    File:
    \texttt{water\_utilities/water\_utilities-dynamic\_properties.xlsx}
  \item
    Sheet: \texttt{water\_price-fixed}
  \item
    Scope: Water Utility
  \item
    Unit: {[} €/house {]}
  \end{itemize}
\item
  Property: \textbf{Water Price Variable Component}

  \begin{itemize}
  \tightlist
  \item
    File:
    \texttt{water\_utilities/water\_utilities-dynamic\_properties.xlsx}
  \item
    Sheet: \texttt{water\_price-variable}
  \item
    Scope: Water Utility
  \item
    Unit: {[} \(\text{€}/m^3\) {]}
  \end{itemize}
\item
  Property: \textbf{Water Price Exchange Rate}

  \begin{itemize}
  \tightlist
  \item
    File:
    \texttt{water\_utilities/water\_utilities-dynamic\_properties.xlsx}
  \item
    Sheet: \texttt{water\_price-selling}
  \item
    Scope: Water Utility
  \item
    Unit: {[} \(\text{€}/m^3\) {]}
  \end{itemize}
\end{itemize}

\hypertarget{masterplan-files}{%
\subsection*{Masterplan Files}\label{masterplan-files}}

The masterplan for each competition stage should be prepared using one
of the template files provided in the supplementary materials. All three
file formats are accepted to provide a trade-off between flexibility and
usability.

The Excel file is the most user-friendly way to describe the solution.
Each lever is described in a separate sheet and requires adding one
entry per row and filling in the columns. The advantage is that most
options can simply be selected using the dropdown menus in the columns.
The downside is that it doesn't offer the flexibility to produce custom
policies for budget allocation
(Section~\ref{sec:policy-budget-allocation}) and NRW interventions
(Section~\ref{sec:policy-nrw-mitigation}).

The YAML format describes the masterplan using the same structure we
introduce for all the levers outlined in
Section~\ref{sec:system-levers}. It allows users to define custom
policies and to populate the file programmatically. However, preparing
it may be somewhat more difficult as it needs to follow the specific
structure shown in Listing~\ref{lst:yaml-format}. The masterplan can be
provided in either JSON or YAML format as long as the structure remains
unchanged.

\begin{codelisting}

\caption{Structure of the masterplan in YAML.}

\hypertarget{lst:yaml-format}{%
\label{lst:yaml-format}}%
\begin{Shaded}
\begin{Highlighting}[]
\FunctionTok{years}\KeywordTok{:}
\AttributeTok{  }\KeywordTok{{-}}\AttributeTok{ }\FunctionTok{year}\KeywordTok{:}\AttributeTok{ YYYY                                                                                  }
\AttributeTok{    }\FunctionTok{national\_policies}\KeywordTok{:}
\CommentTok{      \# Amend here budget allocation and other national policies}
\AttributeTok{    }\FunctionTok{national\_interventions}\KeywordTok{:}
\CommentTok{      \# Install here inter{-}provincial pipes connecting water utilities}
\AttributeTok{    }\FunctionTok{water\_utilities}\KeywordTok{:}
\AttributeTok{      }\KeywordTok{{-}}\AttributeTok{ }\FunctionTok{water\_utility}\KeywordTok{:}\AttributeTok{ WUxx}
\AttributeTok{        }\FunctionTok{policies}\KeywordTok{:}
\CommentTok{        \# Amend here policies for this water utility}
\AttributeTok{        }\FunctionTok{interventions}\KeywordTok{:}
\CommentTok{        \# Report here interventions for this water utility (e.g., pipe installations)}
\end{Highlighting}
\end{Shaded}

\end{codelisting}

\end{document}